\RequirePackage[l2tabu]{nag}
\documentclass{article}

\usepackage[utf8]{inputenc}
\usepackage[T1]{fontenc}

\usepackage{geometry,subfigure}
\usepackage{color}

\usepackage{hyperref}

\usepackage{authblk}
\usepackage{caption}   
\usepackage{color}
\usepackage{graphicx}

\usepackage{amsfonts}
\usepackage{bm}
\usepackage{mathtools}
\usepackage{multirow}

\newcommand{\id}{\mbox d}
\newcommand{\vc}{\bm}
\newcommand{\bnabla}{\bm \nabla}

\title{Are extreme dissipation events predictable in turbulent fluid flows?}

\author[1]{Patrick J. Blonigan} 
\author[2]{Mohammad Farazmand}
\author[2]{Themistoklis P. Sapsis
\thanks{Corresponding author: \href{mailto:sapsis@mit.edu}{sapsis@mit.edu},
Tel: (617) 324-7508, Fax: (617) 253-8689%
}}

\affil[1]{{\small NASA Ames Research Center, Moffett Field, CA 94035, United States}}
\affil[2]{{\small Department of Mechanical Engineering, Massachusetts Institute of Technology,
Cambridge, MA 02139, United States}}

\date{}

\usepackage{amsmath}
\usepackage{amssymb}

\begin{document}

\maketitle
\begin{abstract}
We derive precursors of extreme dissipation events in a turbulent channel flow. Using a recently developed method that combines dynamics and statistics for the underlying attractor, we extract a characteristic state that precedes laminarization events that subsequently lead to extreme dissipation episodes. Our approach utilizes coarse statistical information for the turbulent attractor, in the form of second order statistics, to identify high-likelihood regions in the state space. We then search within this high probability manifold  for the state that leads to the most finite-time growth of the flow kinetic energy. This state has both high probability of occurrence and leads to extreme values of dissipation. We use the alignment between a given turbulent state and this critical state as a precursor for extreme events and demonstrate its favorable properties for prediction of extreme dissipation events. Finally, we analyze the physical relevance of the derived precursor and show its robust character for different Reynolds numbers. Overall, we find that our choice of precursor works well at the Reynolds number it is computed at and at higher Reynolds number flows with similar extreme events. 
\end{abstract}

\section{Introduction}\label{sec:intro}
Turbulent fluid flows have been the most challenging paradigm of chaotic behavior with signatures of persistent and intermittent (i.e. over finite-times and at arbitrary time instants) instabilities leading to nonlinear energy transfers between scales. These nonlinear energy transfers are responsible for both the broad band character of the spectrum but also for the non-Gaussian statistics. More specifically, while non-zero third order statistics are primarily responsible for the persistent non-linear energy transfers (turbulent cascades) and the shape of the spectrum \cite{sapsis_majda_mqg, sapsis_majda_tur}, intermittent events, such as dissipation bursts, are primarily responsible for the heavy tail characteristics \cite{Farazmand2017a,mohamad2016b,mohamad2015}. 

Here we are interested in the formulation of precursors for predicting these extreme events. These are  important in problems related to atmospheric and climate science, fluid-structure interactions, and turbulent fluid flow control, just to mention a few. We present our analysis on a standard configuration of a turbulent fluid flow, namely the channel flow, that exhibits extreme events in the form of large dissipation episodes occurring in random times \cite{Kim_Moin_Moser_1986}. These extreme events rise out of a high-dimensional turbulent attractor essentially without any clear warning. They have the form of a short-time excursion towards laminarization of the flow and a subsequent burst of turbulent kinetic energy which leads to a large dissipation episode pushing the flow away from the laminar state.

Many aspects of these intermittent bursts remain elusive primarily because of the intrinsic high dimensionality of the underlying turbulent attractor that limits the applicability of existing mathematical approaches~\cite{faraz_adjoint,Farazmand2018a}. In particular, extreme events due to their rare character cannot be `seen' effectively by energy-based methods, such as Proper Orthogonal Decomposition (POD). Even if one tries to consider conditional POD modes during extreme events these will not necessarily give the modes related to the triggering of the extreme events, as these do not necessarily obtain high energy, even during the extreme event. 

A different class of methods focus on the spectral analysis of the underlying Koopman operator \cite{rowley09,mezic2013}, and strive 
to extract unstable modes associated with certain observables of the system. Such unstable modes are typically estimated
by Dynamic Mode Decomposition (DMD)~\cite{schmid10,chen12,williams2015}.
This analysis, however, can only detect modes associated with long-term instabilities which
do not seem to explain short-term intermittent events 
observed in turbulent flows~\cite{Farazmand2016,babaee17}.  Other variants, however, such as 
multi-resolution DMD \cite{Kutz2015} have been demonstrated to work well in systems 
with relatively low-dimensional attractors. 

Extreme events in complex dynamical systems have also been analyzed recently using Large Deviation Theory (LDT), e.g. in nonlinear water waves \cite{dematteis2017}. The basic idea is to search the phase space for initial conditions associated with a given magnitude of the objective function (observable of interest) and then from those pick the one with the highest probability of occurrence. However, these efforts have shown to work well in systems where the core of the attractor has Gaussian statistics. For different cases there is no rigorous foundation for LDT to operate. Even in the Gaussian case, the resulting optimization problem has very high dimensionality to be practically solvable for an application like the one considered here  -- see \cite{sapsis2018} for a detailed discussion.

Here we apply a recently developed framework for the discovery of precursors to extreme events \cite{Farazmand2017a}. In contrast to LDT, in the present approach we consider a set of high probability initial conditions, based on a rough approximation of the attractor, and then search within this set for the initial state that has the highest growth for the quantity of interest. An adjoint solver is employed to compute the gradient of the objective function. Because the search is constrained within a low-dimensional (but high-probability) set, the feasibility set, the resulted optimization problem is computationally tractable. The solution provides with an initial state that has high probability to occur and leads to rapid growth of the objective function (in our case the kinetic energy). We discuss the physical relevance of the derived critical state in the context of the turbulent channel flow and use the alignment with this critical state as a scalar precursor for the prediction of future extreme events. We measure the effectiveness of the precursor through direct numerical experiments and examine its robustness over different Reynolds numbers. The success of our approach to an intermittently turbulent channel paves the way for studying transitional flows, such as bypass transition of boundary layers.

This paper is organized as follows. In Section~\ref{sec:prelim} we described the minimal channel flow used in this work
and discuss various aspects of the problem. The optimization problem for discovering the
precursor to extreme events is presented in Section~\ref{sec:optim}. In Section~\ref{sec:pred}
we present a statistical analysis quantifying the predictive power of the optimal precursor. 
Finally, our concluding remarks are presented in Section~\ref{sec:concl}.

\section{Turbulent channel flow - phenomenology}\label{sec:prelim}


\subsection{The minimal flow unit for channel flows}

Turbulent channel flow has been a staple of numerical studies of turbulence for many years \cite{Kim_Moin_Moser_1986}. The chaotic nature of these simulations makes it difficult to analyze local spatiotemporal events and physical mechanisms in them, such as the formation and destruction of individual hairpin vortices in the near-wall region. To isolate these physical mechanisms and others, work has been done to find ``minimal flow units'' for various regions of the channel. Jimenez and Moin~\cite{Jimenez_Moin_1991} found the minimal flow unit for near wall turbulence for several low Reynolds number flows by considering turbulent channel flow simulations with domains that were considerably smaller than conventional channel flow simulations. These smaller domains eliminate larger scale turbulent structures but accurately resolved the near-wall turbulent flow, matching turbulent flow statistics from experiments and prior numerical studies up to 40 wall units in the wall-normal direction. 

Various minimal flow units have been used in a range of different studies because of its isolation of a few physical mechanisms and its relatively low computational cost. Carlson and Lumley used the minimal flow unit to study flow control strategies for turbulent boundary layers \cite{Carlson_Lumley_1996}. Minimal flow units have also been used to study near-wall and log-layer turbulence \cite{Jimenez_Pinelli_1999,Jimenez_Simens_2001,Flores_Jimenez_2010}. These studies use forcing functions to achieve turbulent flows in half-channels, and to selectively damp larger scale flow structures. Recently, near-wall minimal flow units have been used to study and characterize the effects of wall-roughness on wall-bounded flows and to build models of wall roughness effects \cite{Chung_2015,MacDonald_2017}. The near-wall minimal flow unit has also been used to demonstrate shadowing-based adjoint sensitivity analysis \cite{Blonigan_2017}. 

In addition, the near-wall minimal flow unit simulations routinely show highly intermittent behavior which is of interest for the purposes of our study. In certain low Reynolds number simulations, the flow on one wall exhibits turbulent behavior while the other remains laminar. The flow on both walls transitions at seemingly random intervals: the laminar wall would become turbulent, and then the turbulent wall would become laminar. Turbulent flow in the near-wall minimal flow unit is itself intermittent when it occurs. Time series of wall shear stress show that turbulence undergoes a cycle where it proliferates rapidly or ``bursts'', then decays slowly. This observation led to numerous subsequent studies into the intermittent nature of near-wall turbulence using minimal flow units (see Refs~\cite{Jimenez_2012,jimenez2018}, for comprehensive reviews).

\subsection{Flow Solver}

In this study we consider near-wall minimal flow units similar to those considered in Ref. \cite{Jimenez_Moin_1991}. We use a Discontinuous-Galerkin spectral-element method (DGSEM) framework to simulate the minimal flow unit with the compressible Navier-Stokes equations~\cite{Carton_Diosady_2018}. The DGSEM framework has been successfully applied to a range of different flows including channels flows and the near-wall minimal flow unit  \cite{Diosady_Murman_2014,Carton_Murman_2017,Garai_Diosady_Murman_Madavan_2017}. Also, it has an adjoint capability~\cite{Ceze_Diosady_Murman_2016} that has been validated for the near-wall minimal flow unit in Ref. \cite{Blonigan_2017}. A detailed description of the discretization and implementation of this solver is available in Refs. \cite{Diosady_Murman_2013} and \cite{Diosady_Murman_2015}.

We run a direct numerical simulation (DNS) of the channel flow with the compressible Navier-Stokes equations with a constant forcing in the axial direction to drive the channel. Since the Mach number is low, the effective governing equations reduce to the incompressible 
Navier--Stokes equations
\begin{subequations}\label{e:NSincompresible}
\begin{equation}
\bnabla\cdot\vc u=0,
\end{equation}
\begin{equation}
\partial_t \vc u +\vc u\cdot \bnabla\vc u = \frac{f_0}{\rho} \vc e_1 -\frac{1}{\rho}\bnabla p +\nu \Delta \vc u,
\end{equation}
\begin{equation}
\vc u(\vc x,0)=\vc u_0(\vc x)
\end{equation}
\end{subequations}
where $\vc u = (u,v,w)$ denotes the three-dimensional velocity field with streamwise component $u$, 
wall-normal component $v$ and spanwise component $w$. The constant forcing in the
streamwise direction is denoted by $f_0$ and $\vc e_1=(1,0,0)$ is the unit vector in the streamwise direction. The boundary conditions are no-slip at the walls so that $\vc u(x,\pm \delta,z,t)=0$, 
and periodic in the spanwise and streamwise directions. Here, the channel height in the wall-normal direction is $2\delta$.

\subsection{Numerical experiment set-up}

The case we consider has a domain size of $\pi\delta \times 2\delta \times 0.34\pi\delta$ in the streamwise, wall-normal, and spanwise directions, respectively.  The channel half-height $\delta$ was set to $1.0$. The flow considered is at Reynolds number $Re=2200$ (corresponding to the friction Reynolds number $Re_{\tau}=110$) unless stated otherwise. The Reynolds number $Re$ is defined as
$Re = \rho U \delta/\mu$
where $\rho$ is the fluid density, $U$ is the centerline velocity of a laminar flow with the same mass flow rate (as in Ref. \cite{Jimenez_Moin_1991}), and $\mu$ is the dynamic viscosity. As in Ref. \cite{Blonigan_2017}, $U$ was chosen so that the Mach number of the flow was under $0.3$ and therefore the flow is nearly incompressible. The friction Reynolds number $Re_{\tau}$ is defined as 
$Re_{\tau} = \rho u_{\tau} \delta/\mu$
where $u_{\tau}=\sqrt{\tau_w/\rho}$ is the friction velocity and $\tau_w$ is the average shear stress at the wall. 
The channel constant forcing $f_0$ is set to balance the mean shear stress of the walls, so it is set by the choice of Reynolds number $Re$, and the friction Reynolds number $Re_{\tau}$ as follows
\begin{equation}
f_0 = \frac{\tau_w}{\delta} = \frac{Re_{\tau}^2}{Re} \rho U
\end{equation}

The domain is discretized with a $4 \times 16 \times 2$ mesh with 8th order spatial elements, resulting in a $32 \times 128 \times 16$ distribution of nodes (65536 total), similar to the mesh used in Ref. \cite{Jimenez_Moin_1991}. The choice of $Re_{\tau}=110.0$ results in grid resolutions of $\Delta x^+ \approx 11$ and $\Delta z^+ \approx 7$ wall units per node, where
$x^+ = u_{\tau} x/\nu$ , $y^+ = u_{\tau} y/\nu$, and $z^+ = u_{\tau} z/\nu$.

The wall-normal spacing corresponds to $\Delta y^+ \approx 0.6$ for the nodes closest to the walls, which ensures that the simulations are well resolved. We used space-time elements that were 4th order in time and the time slab (temporal element) was $\Delta t =0.05t_e$, where
$t_e = \delta/U$ denotes the eddy turnover time, the time scale associated with the largest possible eddy in the channel.

%

The DGSEM flow solver used here has been validated for the minimal flow unit at $Re=3000$ \cite{Blonigan_2017} and we carried out a similar validation study for the $Re=2200$ case. Note that the statistics at $Re=2200$ were only computed over time intervals when both walls had turbulent flow, as was done in Ref. \cite{Jimenez_Moin_1991} for low Reynolds number cases. This was necessary because the $Re=2200$ flow exhibited intermittent behavior similar to that observed in previous studies of minimal flow units. 

\subsection{Extreme events}\label{sec:bursts}

\begin{figure}
\centering
\subfigure[\label{f:Re2200_ke}]{\includegraphics[width=.48\textwidth]{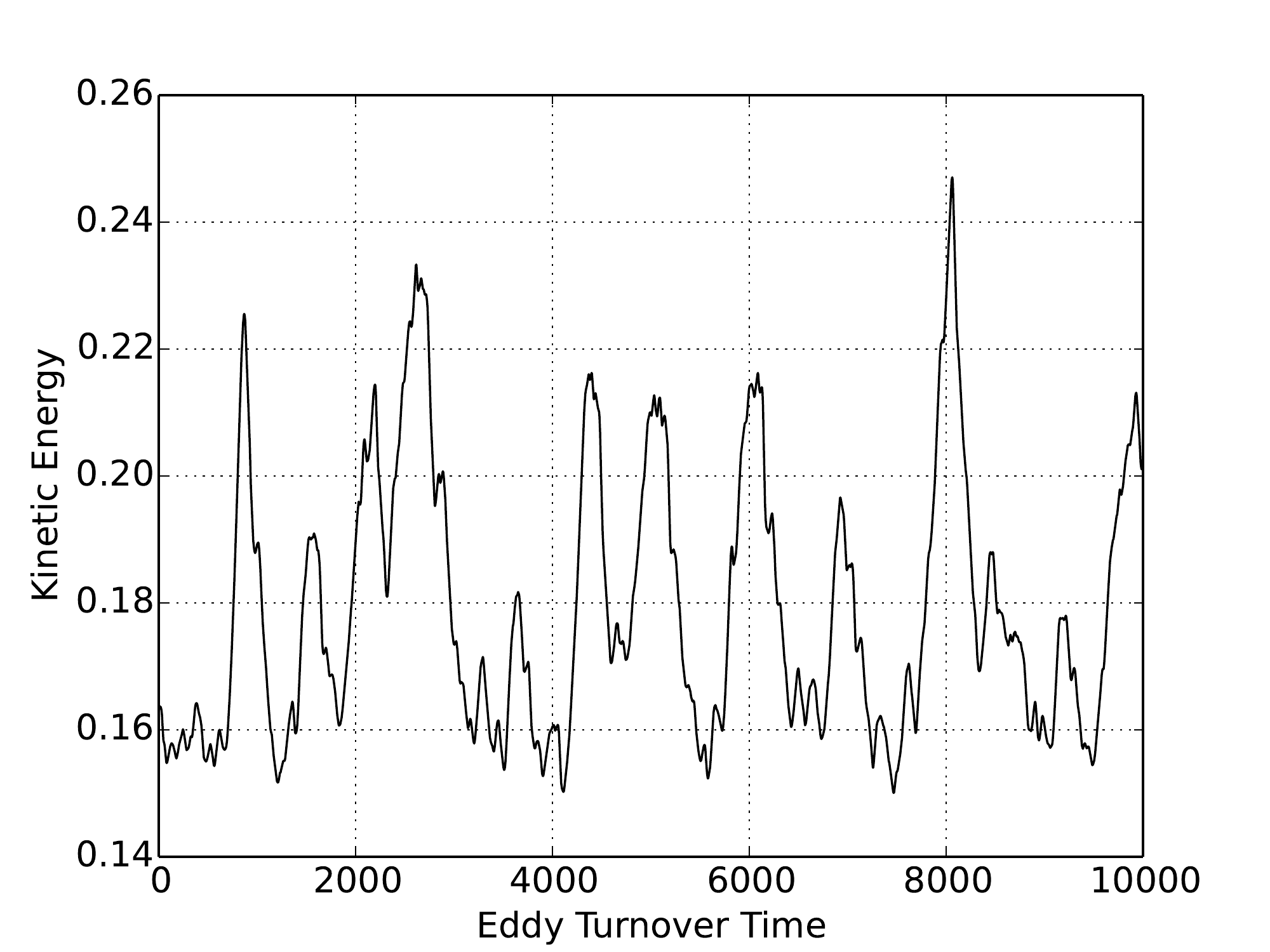}}
\subfigure[\label{f:Re2200_diss}]{\includegraphics[width=.48\textwidth]{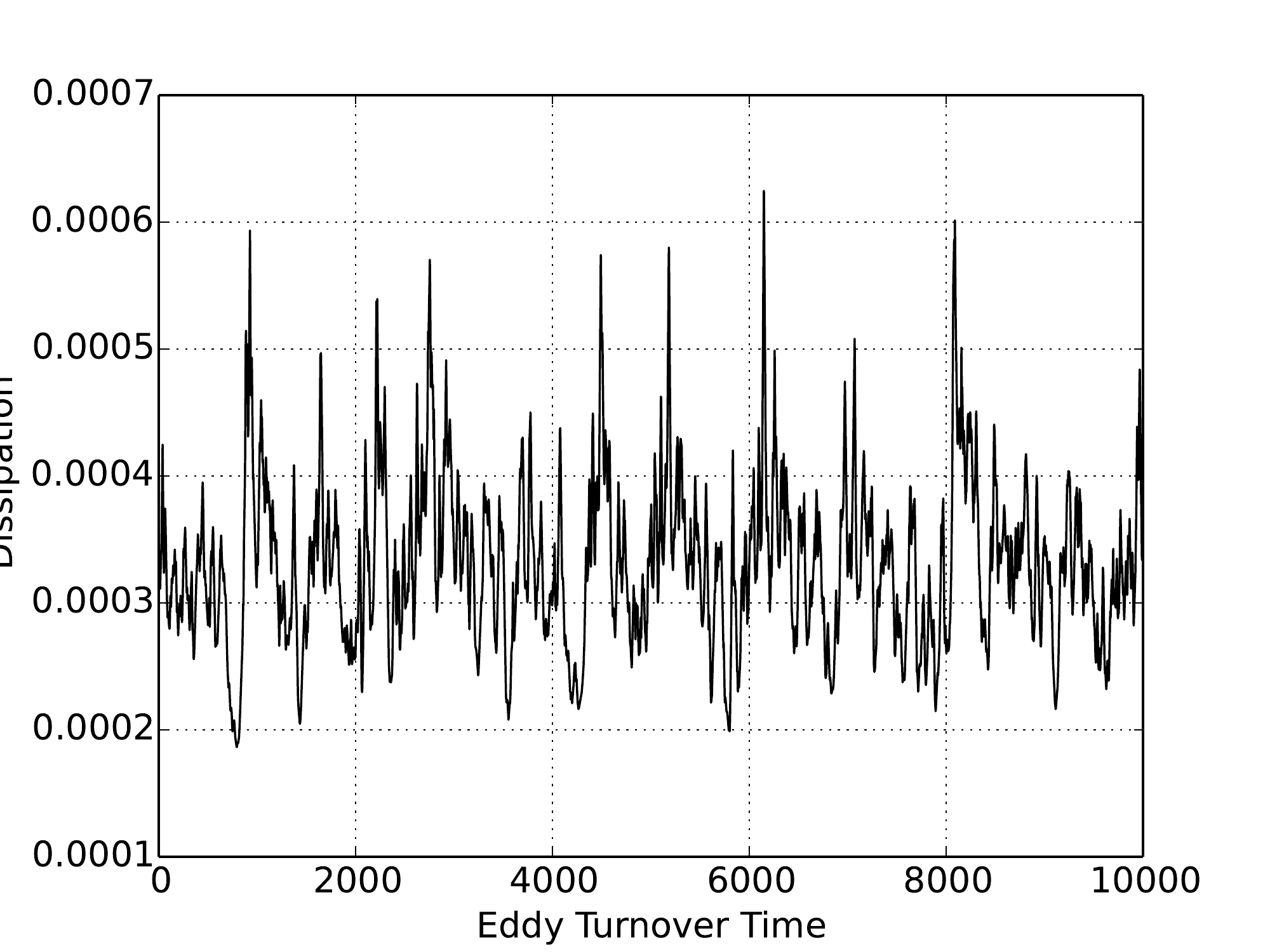}}
\caption{Time evolution of the kinetic energy (a) and the energy dissipation rate (b) for the near-wall minimal flow unit at $Re=2200$ }
\label{f:Re2200_E_Z_hist}
\end{figure}

The intermittent behavior of the flow at $Re=2200$ can be seen in Fig. \ref{f:Re2200_E_Z_hist} from the spikes in kinetic energy $E(t)$ and dissipation $Z(t)$. We define the kinetic energy,
\begin{equation}
E(t) = \iiint_{\Omega} \rho \vc u \cdot \vc u \, \id x \, \id y \, \id z,
\label{e:ke_def}
\end{equation}
where $\Omega$ is the flow domain. Energy dissipation rate $Z(t)$ is defined as
\begin{equation}
Z(t) = \iiint_{\Omega} \text{tr}(\boldsymbol{\tau} \nabla \vc u ) \, \id x \, \id y \, \id z,
\label{e:diss_def}
\end{equation}
where $\boldsymbol{\tau}$ denotes the stress tensor, defined as 
$\vc \tau = \mu(\nabla \vc u + \nabla \vc u^\top)$
for an incompressible flow.

Fig. \ref{f:Re2200_E_Z_hist} shows that large increases in $E(t)$ are followed by spikes in $Z(t)$ and a subsequent decrease in $E(t)$. These large spikes in kinetic energy occur when there is laminar flow near one wall, and nearly laminar flow near the other wall. Laminar flows correspond to higher kinetic energy $E(t)$ because the channel is driven by a constant body force in the axial direction. This body force acts as a fixed axial pressure gradient. The body force and wall shear stress balance one another when the flow is in an equilibrium state where we have
\begin{equation}
\mu \frac{\partial \overline{u}}{\partial n}  = \delta \frac{\partial \overline{p}}{\partial x},
\end{equation}
where $n$ is the wall-normal direction. For a given centerline velocity, a laminar flow will exert less shear stress $\mu \frac{\partial \overline{u}}{\partial n}$ on the walls than a turbulent flow, so for a given wall shear stress $\mu \frac{\partial \overline{u}}{\partial n}$ the laminar flow will have a larger centerline velocity than a turbulent flow. 

%

\begin{figure}
\centering
\includegraphics[width=.6\textwidth]{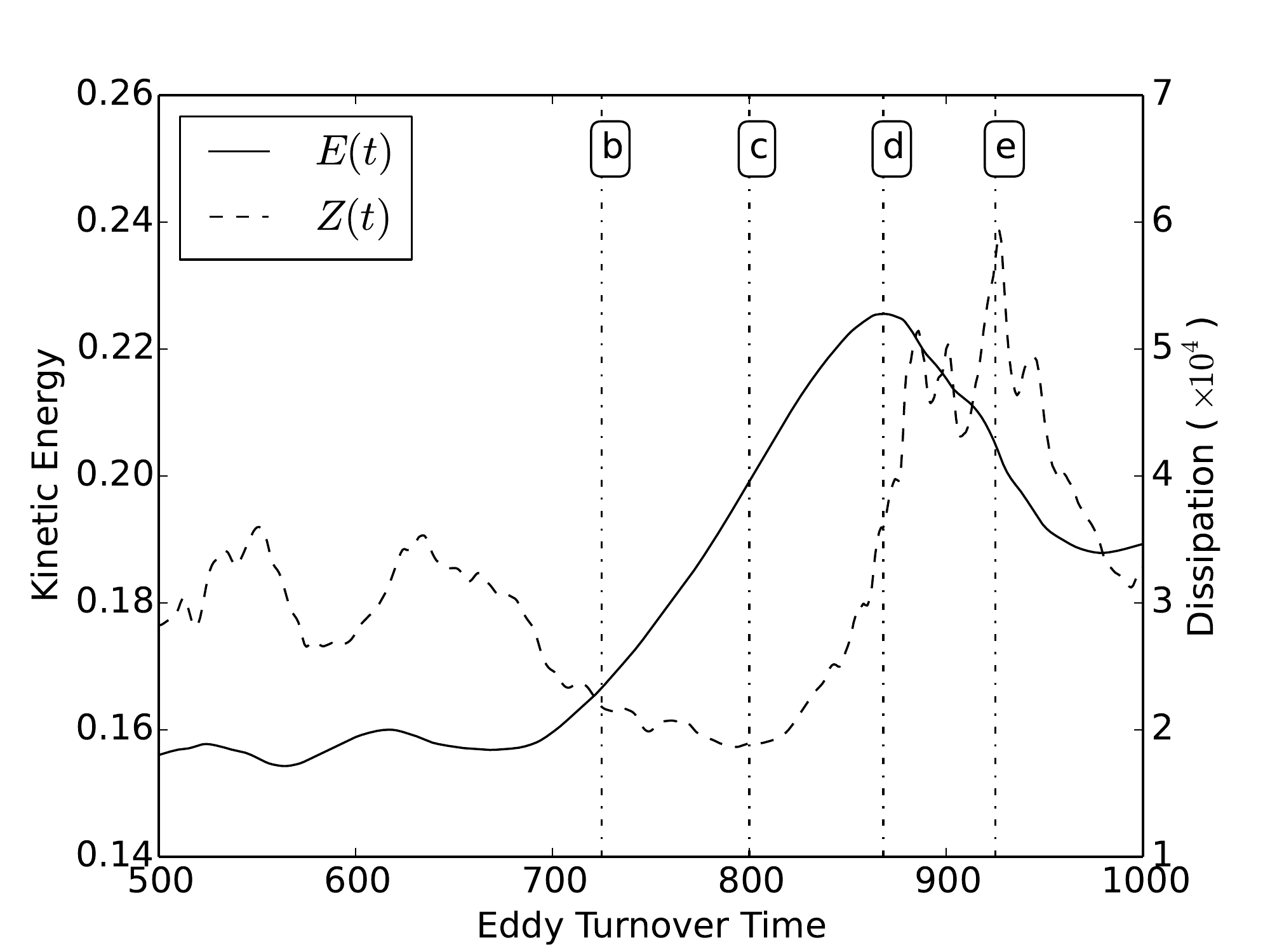}
\caption{Time evolution of $E(t)$ and $Z(t)$ during the first laminarization event shown in Fig. \ref{f:Re2200_E_Z_hist}. The vertical lines indicate the times that snapshots in Fig. \ref{f:extreme_event_snapshots} correspond to. The first and last snapshot correspond to the start and end of the time horizon shown above.  }
\label{f:ee_ke_diss}
\end{figure}

\begin{figure}
\centering
\subfigure[$t=500t_e$\label{f:snapshot1}]{\includegraphics[width=.3\textwidth]{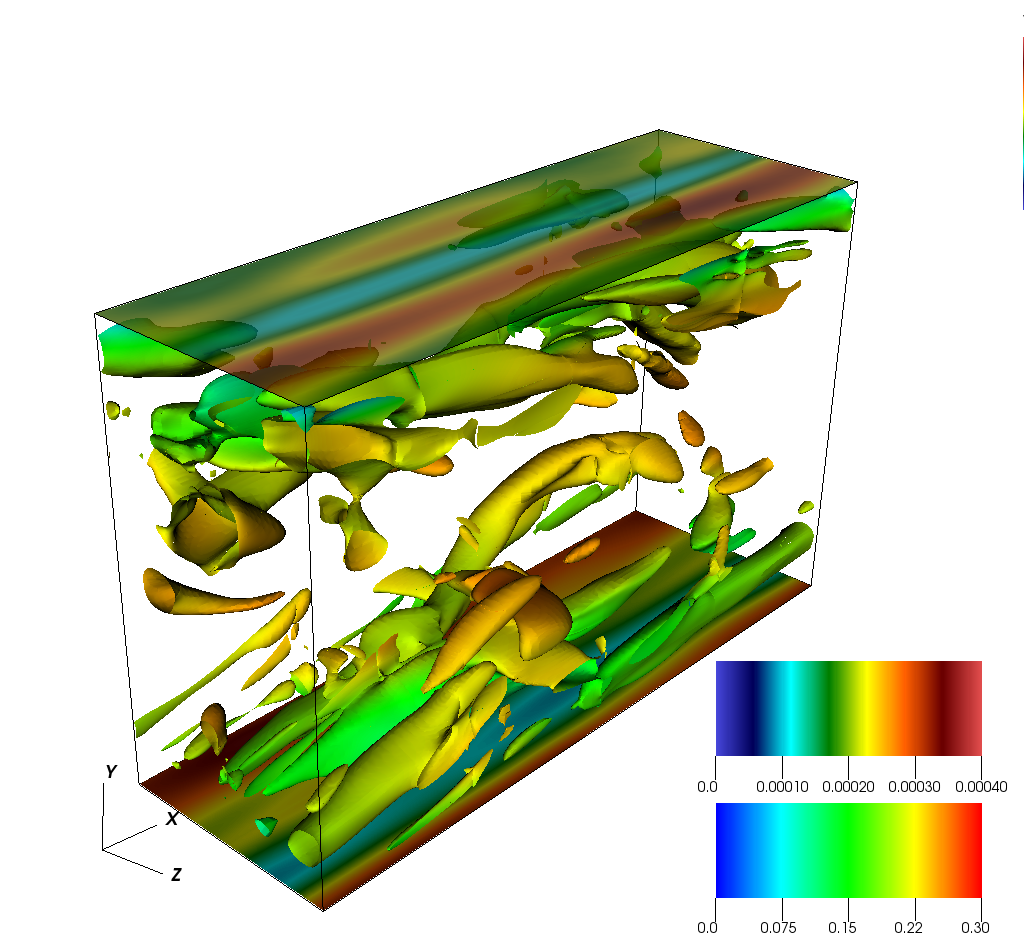}}
\subfigure[$t=725t_e$\label{f:snapshot2}]{\includegraphics[width=.3\textwidth]{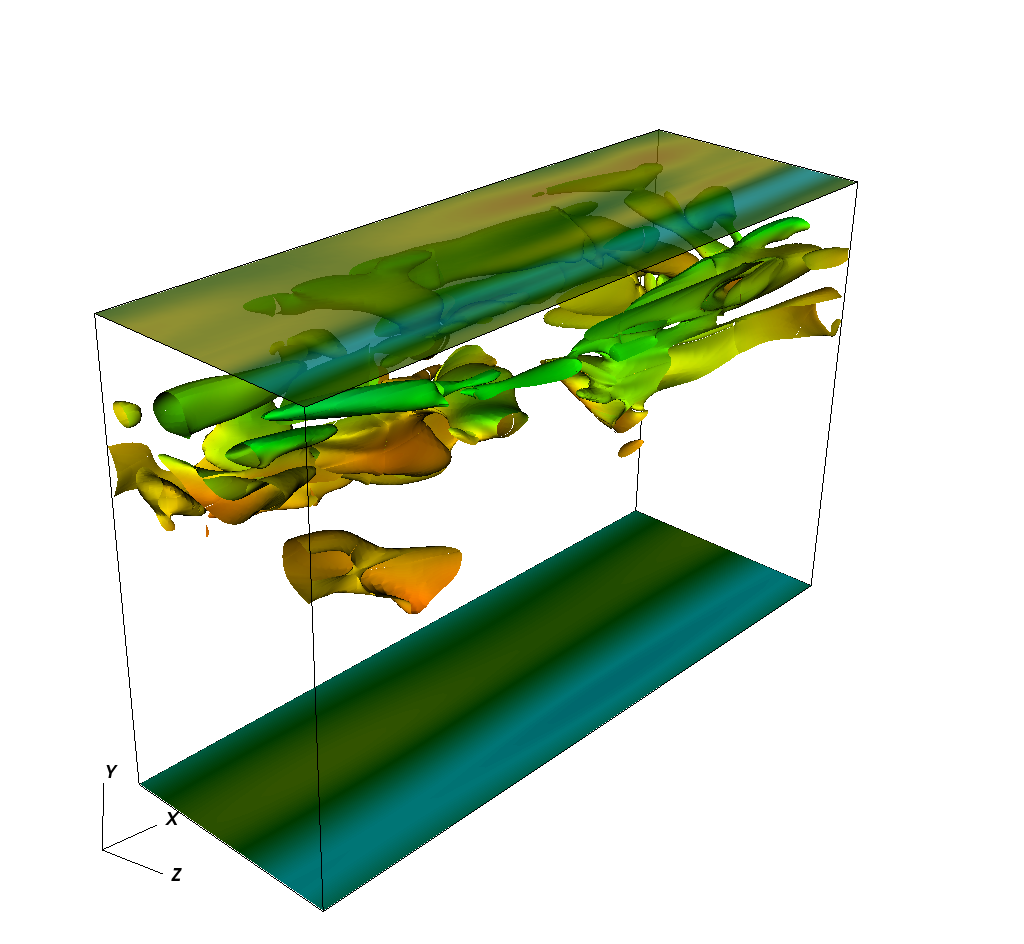}}
\subfigure[$t=800t_e$\label{f:snapshot3}]{\includegraphics[width=.3\textwidth]{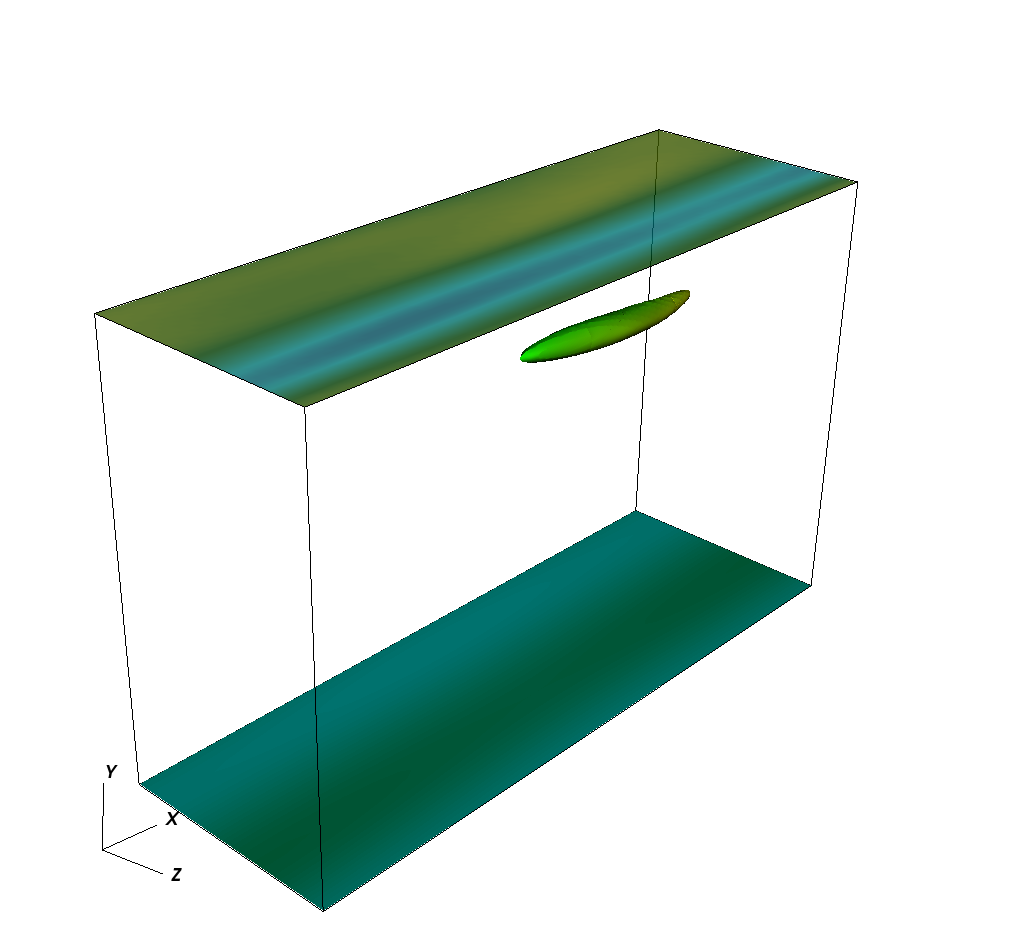}}
\subfigure[$t=868t_e$\label{f:snapshot4}]{\includegraphics[width=.3\textwidth]{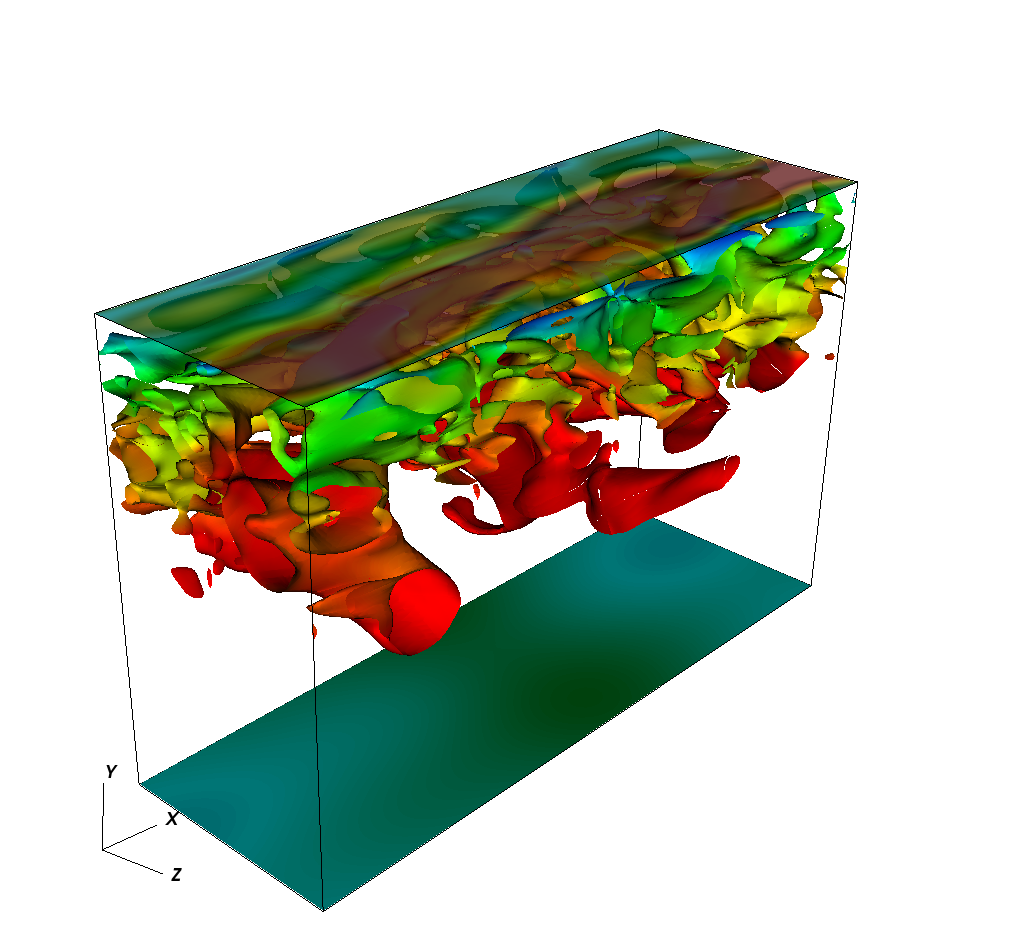}}
\subfigure[$t=925t_e$\label{f:snapshot5}]{\includegraphics[width=.3\textwidth]{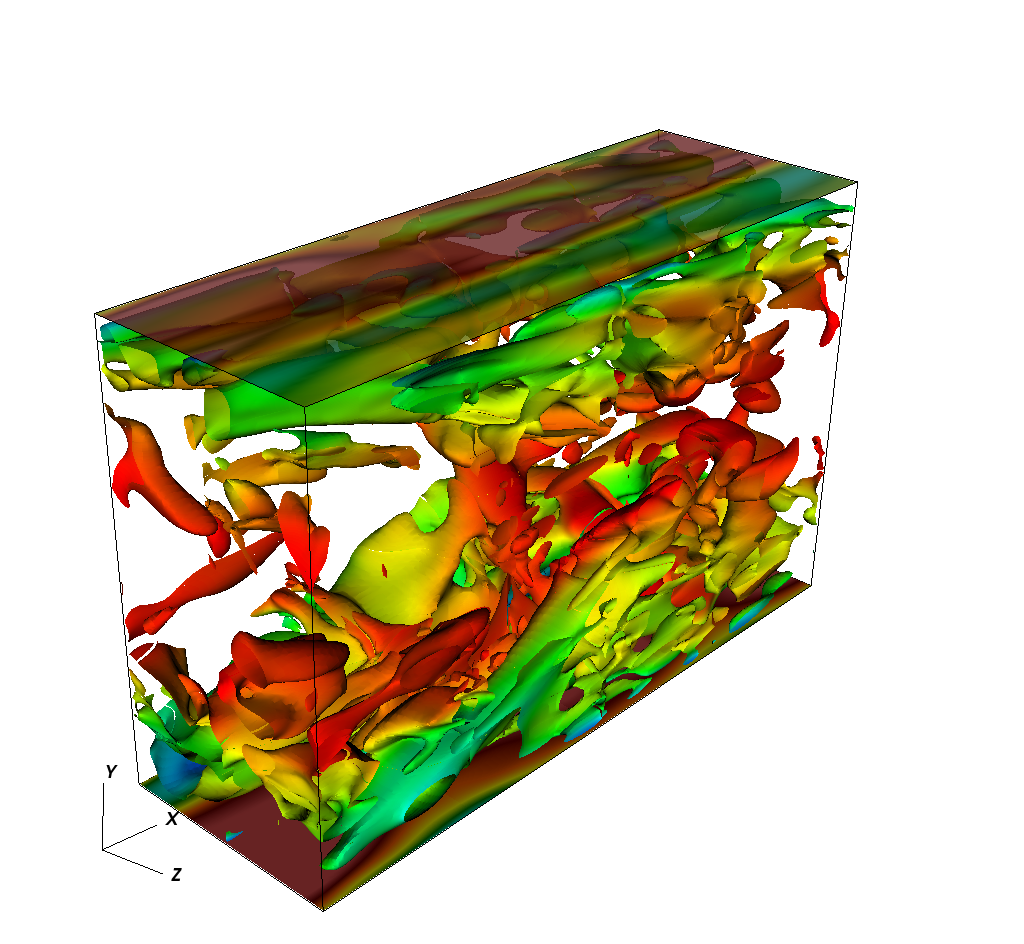}}
\subfigure[$t=1000t_e$\label{f:snapshot6}]{\includegraphics[width=.3\textwidth]{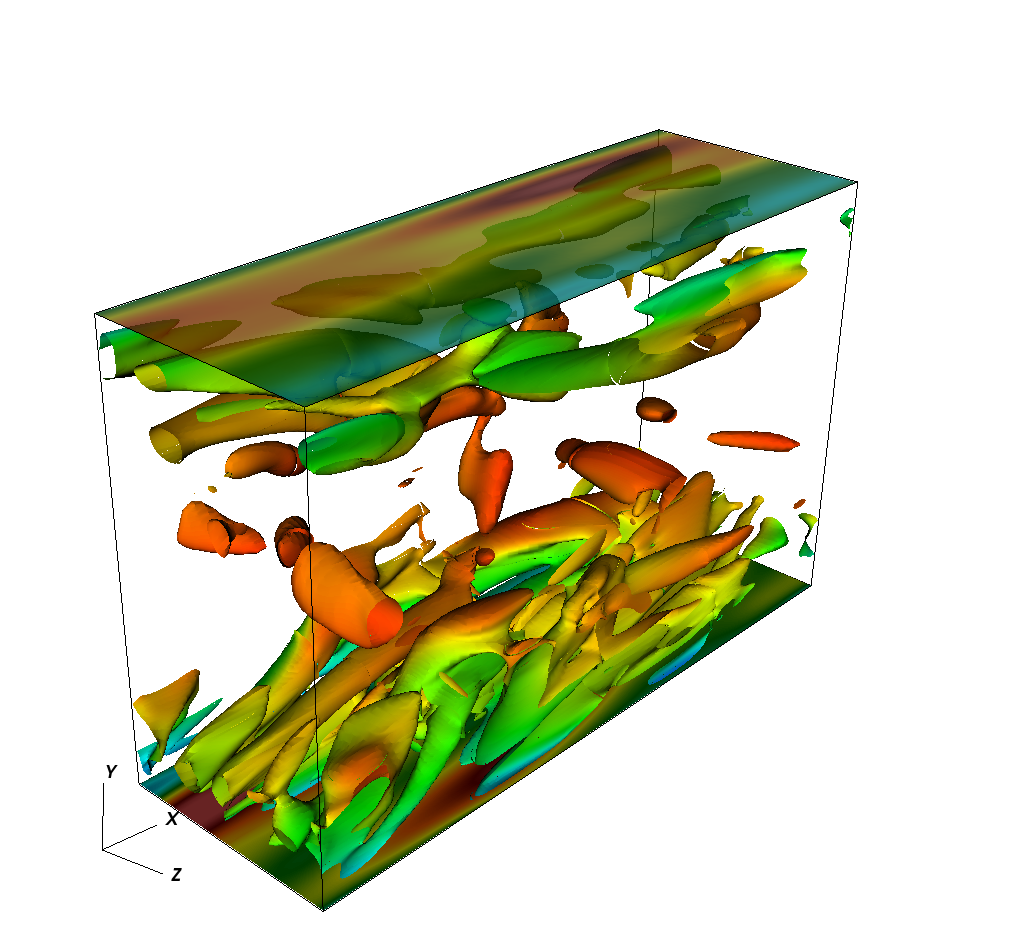}}
\caption{Snaphots of Q-criterion colored by axial velocity and wall shear stress. }
\label{f:extreme_event_snapshots}
\end{figure}

Therefore, the large spikes in $E(t)$ are the result of a flow laminarization event. Of course, the flow never completely reaches the laminar
state, though it gets very close it.
Figs. \ref{f:ee_ke_diss} and \ref{f:extreme_event_snapshots} show an example of a typical laminarization event where
the flow undergoes the following stages.
\begin{enumerate}
\item The flow on the bottom wall becomes laminar (Fig. \ref{f:snapshot2}). 
\item The flow on the top wall becomes nearly laminar (Fig. \ref{f:snapshot3}).
\item As the entire channel becomes nearly laminar, the axial velocity increases. 
\item The higher velocities make the effective Reynolds number of the flow larger. This increases the likelihood of turbulent burst occuring since the flow is less stable to perturbations at a larger Reynolds number. 
\item A turbulent burst occurs on the top wall, which causes $Z(t)$ to increase rapidly (Fig. \ref{f:snapshot4}).
\item The bottom wall transitions to turbulence (Fig. \ref{f:snapshot5}).
\item The turbulent flow on both walls causes $E(t)$ to decrease as it returns to the equilibrium mean turbulent flow profile (Fig. \ref{f:snapshot6}). 
\end{enumerate}

These flow laminarization events, and the resulting bursts of energy 
dissipation rate, are the extreme events we will consider in this paper.

\section{Optimal states for extreme events}\label{sec:optim}
We now describe the constrained optimization
problem whose solutions determine the most likely 
triggers of extreme events. The method is presented in detail in 
Ref.~\cite{Farazmand2017a} and is reviewed here for 
completeness. We describe the optimization problem in the context of the channel flow, 
outline the numerical method for obtaining its solutions and present our numerical 
results.  

\subsection{High-Likelihood triggers of extreme events}

We seek initial states $\vc u_0$ that after a given integration time $\tau$
realize the largest possible energy growth. More precisely, 
we seek initial states $\vc u_0$ such that $E(\vc u(\tau)) - E(\vc u_0)$
is maximized.
This is a PDE-constrained optimization problem,
since the velocity field $\vc u(t)$ is required to satisfy the channel flow~\eqref{e:NSincompresible}.

In addition to this PDE constraint, we also enforce a feasibility constraint, by requiring the initial state $\vc u_0$
to belong to the system attractor. This second constraint is essential in order to guarantee that 
the optimal solution is probabilistically relevant. As in many dissipative PDEs, the channel flow has an attractor towards which the solutions converge asymptotically after an initial transient. We are interested in the self-sustained and recurrent extreme events on this attractor as opposed
to transient extreme events off the attractor that may occur but are not recurrent. In order to prevent the optimizer from considering such transient 
events, we enforce a feasibility constraint which is further elucidated in section~\ref{sec:pod}.

With this prelude, the optimization problem can be formulated as
\begin{subequations}~\label{eq:optim}
\begin{equation}\label{eq:max}
\max_{\vc u_0\in\mathcal U} \big[E(\vc u(\tau))-E(\vc u_0) \big],
\end{equation}
\begin{equation}\label{eq:const_1}
\vc u(\tau)\ \mbox{satisfies equation~\eqref{e:NSincompresible} with}\ \vc u(0)=\vc u_0,
\end{equation}
\begin{equation}\label{eq:const_2}
\vc u_0\in\mathcal A\subset\mathcal U,
\end{equation}
\end{subequations}
where $\tau>0$ is a prescribed time, related to the growth time scale of a typical extreme event. The constraint~\eqref{eq:const_1} implies that $\vc u(\tau)$
is a solution of the channel flow. Constraint~\eqref{eq:const_2} ensures that the 
optimizer belongs to the attractor $\mathcal A$ and is therefore probabilistically relevant. 
In the next section, we describe the method used to approximate the attractor $\mathcal A$. 
\begin{figure}
\centering
\includegraphics[width=.5\textwidth]{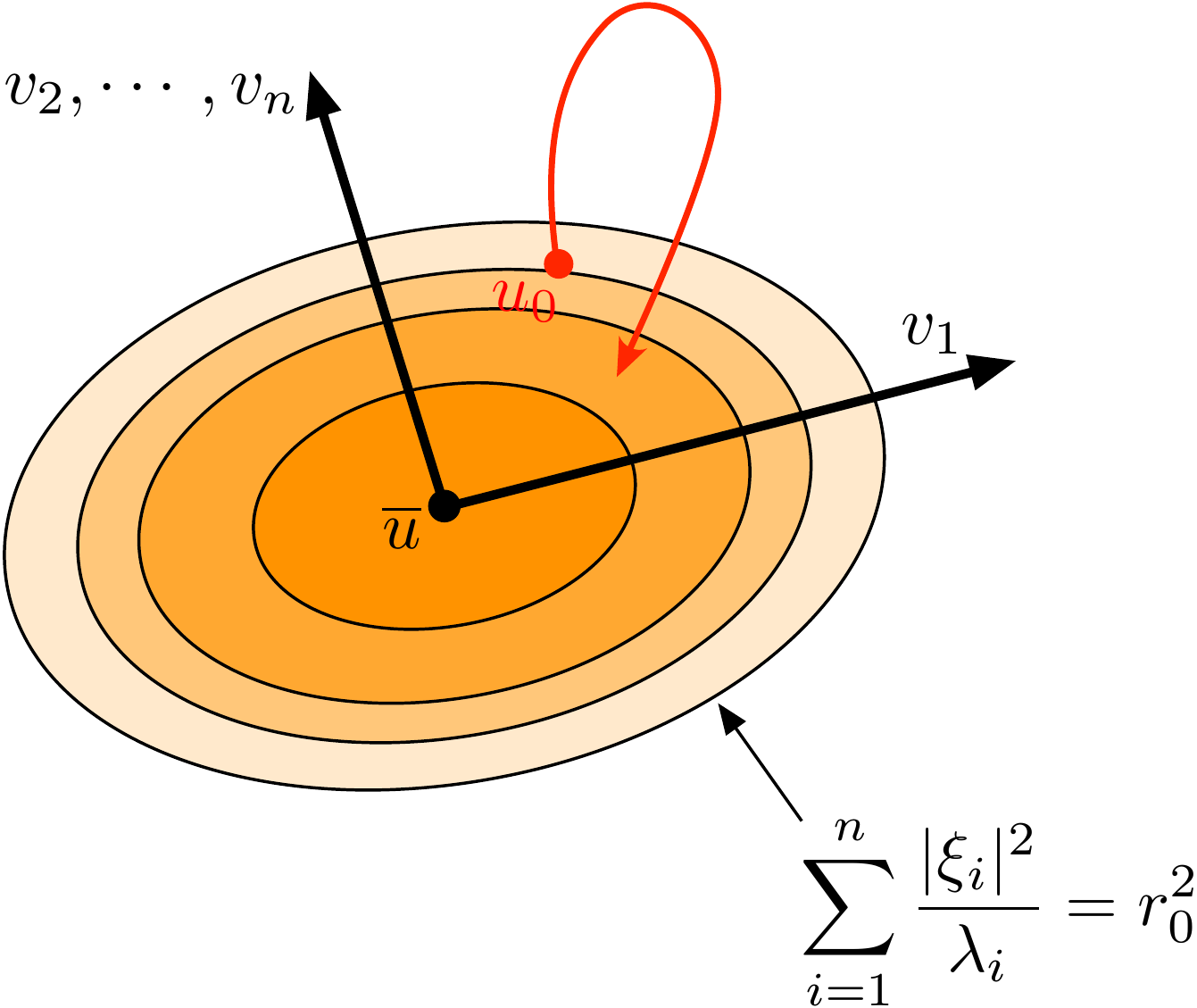}
\caption{A sketch of the principle component analysis of the turbulent data. The
attractor is approximated as an ellipsoid in the subspace spanned by the POD modes $\{\vc v_1,\cdots,\vc v_n\}$.
The origin is the mean flow $\overline{\vc u}$.
}
\label{fig:ellipsoid}
\end{figure}

\subsection{Feasibility constraint and proper orthogonal decomposition}\label{sec:pod}

The attractors of dissipative dynamical systems are often very complex sets.
Their estimation has been the subject of many studies (see e.g. ~\cite{Holmes_et_al96}). Here, we approximate the attractors
via Proper Orthogonal Decomposition (POD) of longterm simulations of the channel flow
(this method is also known as the Principal Component Analysis). 
The POD approximation assumes that the attractor has a Gaussian
distribution with mean $\overline{\vc u}(\vc x)$ and covariance matrix $\vc C(\vc x,\vc x')$
where 
\begin{subequations}\label{eq:pod}
\begin{equation}\label{eq:pod1}
\overline{\vc u}(\vc x) = \lim_{T\to\infty}\frac{1}{T}\int_{0}^T\vc u(\vc x,t)\id t,
\end{equation}
\begin{equation}\label{eq:pod2}
\vc C(\vc x,\vc x')=\lim_{T\to\infty}\frac{1}{T}\int_{0}^T 
\left(\vc u(\vc x,t)-\overline{\vc u}(\vc x)\right)\otimes
\left(\vc u(\vc x',t)-\overline{\vc u}(\vc x')\right)\id t.
\end{equation}
\end{subequations}
Let the vector fields $\vc v_i:\Omega \to \mathbb R^3$ denote the eigenfunctions of the
covariance tensor, so that
\begin{equation}\label{eq:pod_eig}
\int_\Omega \vc C(\vc x,\vc x')\vc v_i(\vc x')\id \vc x' = \lambda_i \vc v_i(\vc x),\quad i\in\mathbb N
\end{equation}
where $\lambda_i\in\mathbb R$ are the corresponding eigenvalues. The eigenvectors are
ordered such that $\lambda_1\geq \lambda_2\geq \cdots$. 
Since the 
covariance tensor is symmetric and positive definite, the eigenvalues are real-valued and non-negative, and furthermore the eigenfunctions
are orthogonal with respect to the $L^2$ inner product, 
i.e., $\langle\vc v_i,\vc v_j\rangle_{L^2(\Omega)}=\delta_{ij}$.
We refer to the eigenfunctions $\vc v_i$ as the POD modes.

In the POD approximation, any state on the attractor is approximated as 
\begin{equation}
\vc u(\vc x,t) = \overline{\vc u}(\vc x) +\sum_{i=1}^{n} \xi_i(t)\vc v_i(\vc x),
\end{equation}
which is a finite-dimensional truncation to the first $n$ POD modes. Each component of the
vector $\vc\xi = (\xi_1,\cdots,\xi_n)\in\mathbb R^n$ is given by 
$\xi_i(t) =\langle \vc u(t)-\overline{\vc u},\vc v_i\rangle_{L^2(\Omega)}$.

With this POD approximation of the attractor, the optimization problem~\eqref{eq:optim}
can be written more explicitly as
\begin{subequations}~\label{eq:optim_pod}
\begin{equation}\label{eq:max_pod}
\max_{\vc\xi\in\mathbb R^n} \big[E(\vc u(\tau))-E(\vc u_0) \big],
\end{equation}
\begin{equation}\label{eq:const1_pod}
\vc u(\vc x,\tau)\ \mbox{satisfies equation~\eqref{e:NSincompresible} with}\ \vc u(\vc x,0)=\vc u_0(\vc x),
\end{equation}
\begin{equation}\label{eq:const2_pod}
\vc u_0(\vc x)= \overline{\vc u}(\vc x) +\sum_{i=1}^{n} \xi_i\vc v_i(\vc x),
\end{equation}
\begin{equation}\label{eq:const3_pod}
\sum_{i=1}^n\frac{\xi_i^2}{\lambda_i}\leq r_0^2
\end{equation}
\end{subequations}
where $r_0\in\mathbb R$ is a prescribed parameter that is set equal to 1.0 in this study. Note that the form of the constraint essentially restricts our optimization within states that have the highest 
probability, given second-order statistics for the attractor. Constraint~\eqref{eq:const2_pod} enforces that 
the mean-zero initial condition $\vc u_0-\overline{\vc u}$ belongs to the subspace spanned by the first $n$ POD modes. 
Constraint~\eqref{eq:const3_pod}, which describes an ellipsoid in the subspace 
$\mbox{span} \{\vc v_1,\cdots,\vc v_n\}$, ensures that the initial conditions are not two far from the 
mean flow $\overline{\vc u}$ (see Fig.~\ref{fig:ellipsoid}).

\begin{figure}
\centering
\subfigure[]{\includegraphics[width=.48\textwidth]{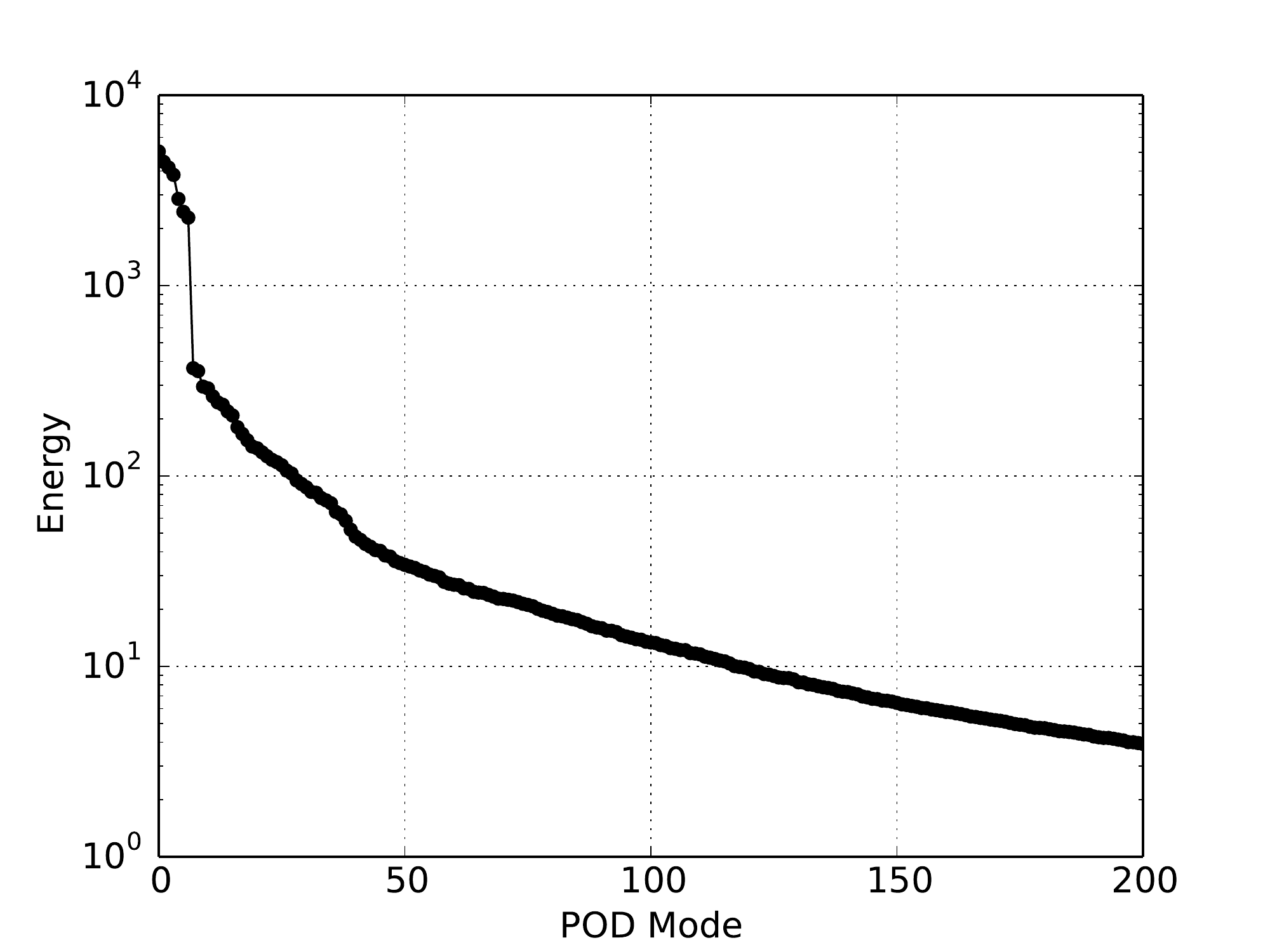}}
\subfigure[]{\includegraphics[width=.48\textwidth]{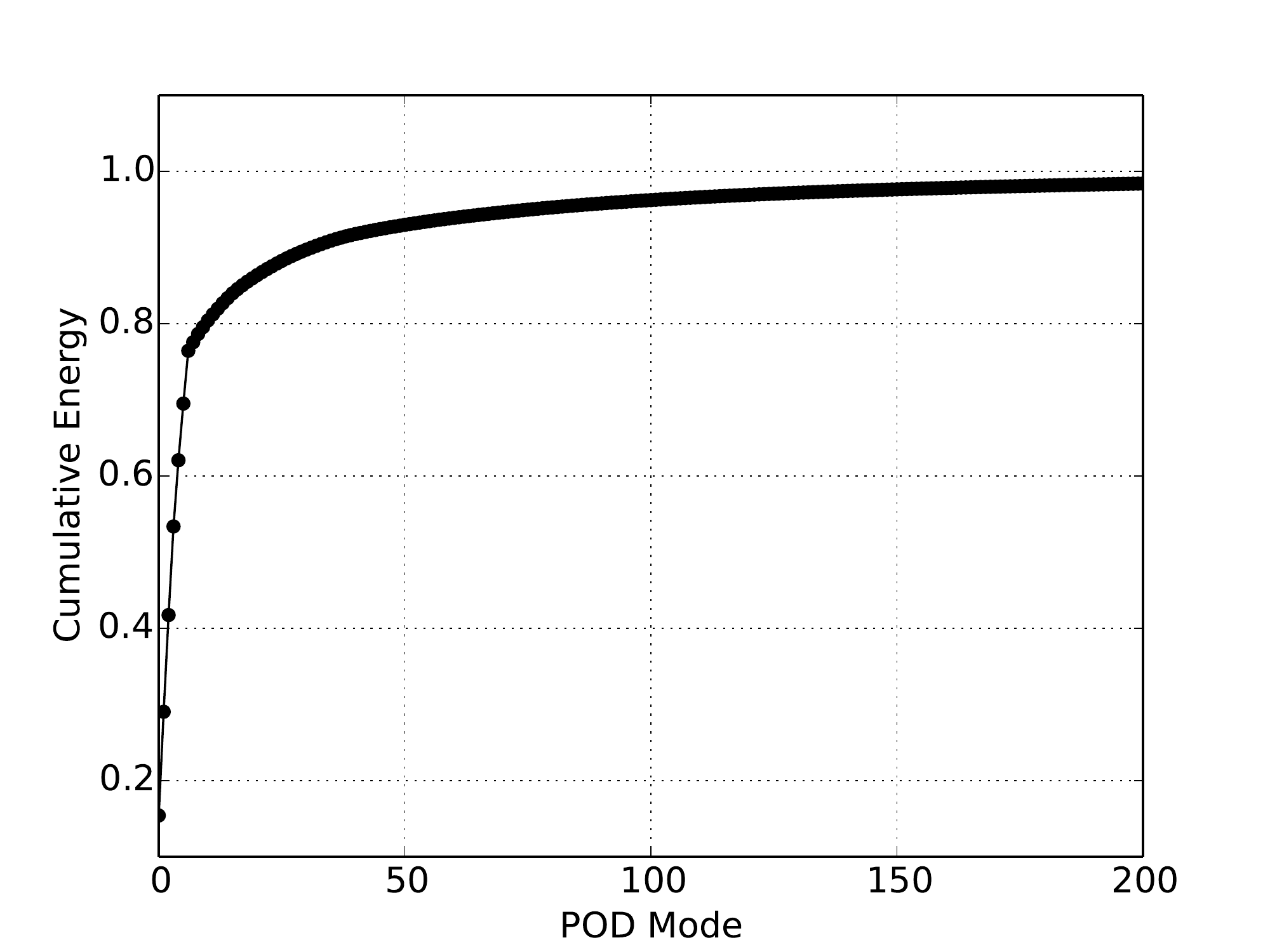}}
\caption{The energy content of the POD modes.
(a) Energy content of each POD mode. 
(b) Cumulative energy content of the POD modes. 
POD was computed with $1000$ snapshots at $Re=2200$. }
\label{f:pod_modes_energy}
\end{figure}

Although the initial condition $\vc u_0$ is 
constrained to the subspace spanned by the first $n$ POD modes, the final state $\vc u(\tau)$
may not belong to this subspace. This is because the POD decomposition is only an approximation 
of the attractor, which represents initial states for our analysis, and not the exact invariant attractor. However, we take $n$ large enough
so that only an insignificant fraction of the energy content of the states on the attractor are neglected. 
More precisely, in the following, we set $n=50$ so that the truncation of the turbulent states to 
these POD modes contain at least $90\%$ of the kinetic energy, as shown in Fig. \ref{f:pod_modes_energy}. 

\begin{figure}[h]
\centering
\includegraphics[width=.5\textwidth]{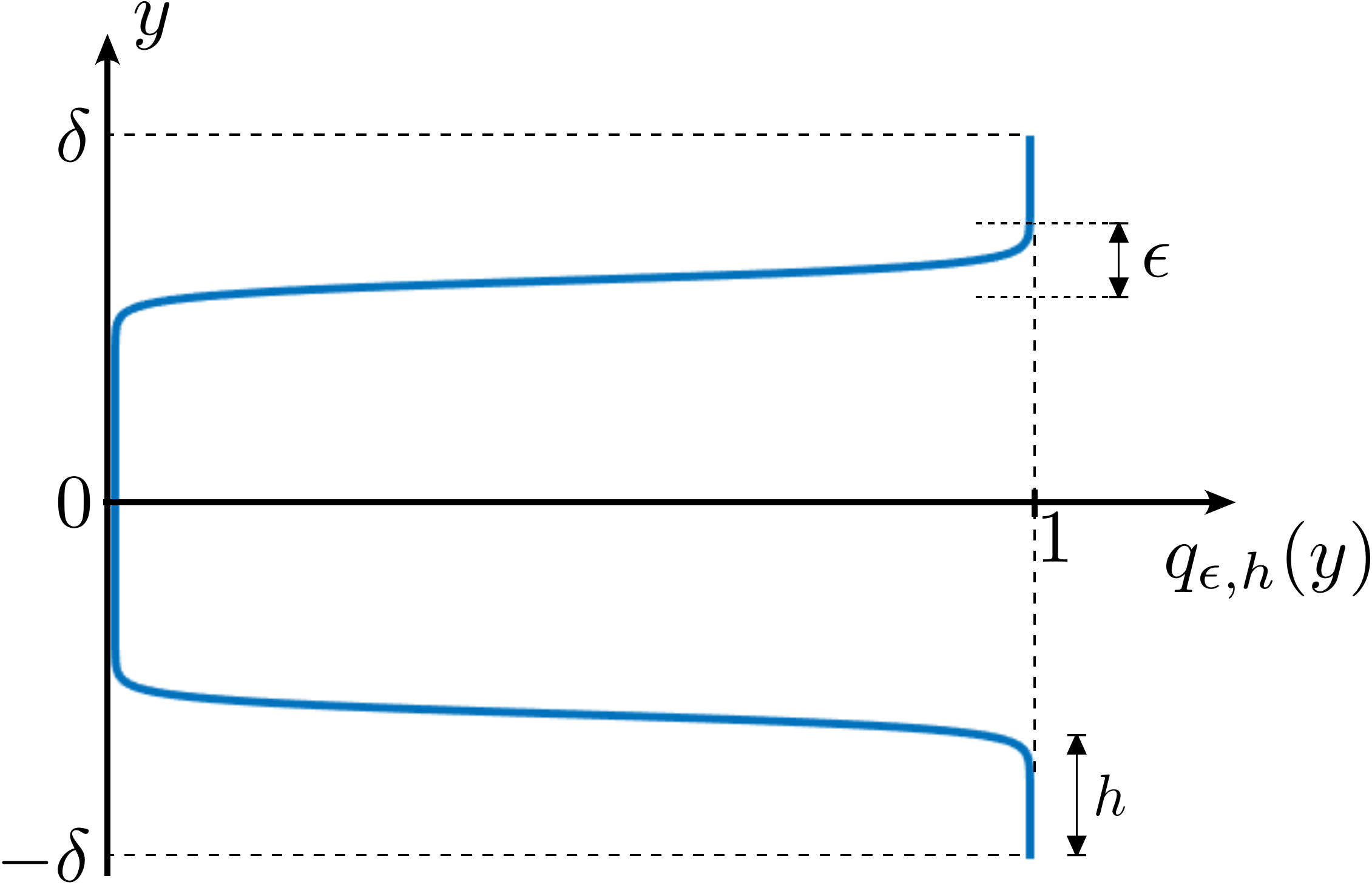}
\caption{A sketch of the weight function $q_{\epsilon,h}$ defined in~\eqref{eq:weight}.}
\label{fig:q}
\end{figure}

As discussed in section~\ref{sec:bursts}, previous studies of the minimal flow unit have shown that intermittent bursts of the flow
originate from the near-wall activities. In light of these observations,
we make an important modification to the computation of the POD modes by 
multiplying the zero-mean velocity fields by a weight function that emphasizes the 
near-wall contribution of the flow. 
More precisely, we compute the weighted velocity fields
\begin{equation}
\vc u_{\epsilon,h}(\vc x,t)=\left(\vc u(\vc x,t)-\overline{\vc u}(\vc x)\right)q_{\epsilon,h}(y),
\end{equation} 
by multiply the original velocity fields $\vc u$ (after removing the mean $\overline{\vc u}$) with the weight function 
\begin{equation}
q_{\epsilon,h}(y)= \frac{1}{2}\left\{
2+\tanh\left[\frac{1}{\epsilon}\big(y-(\delta-h)\big) \right]  
-\tanh\left[\frac{1}{\epsilon}\big(y+(\delta-h)\big) \right]
 \right\}.
\label{eq:weight}
\end{equation}
For $\epsilon\ll h \ll \delta$, the weight function $q_{\epsilon,h}$ vanishes near the 
center of the channel and approaches unity near the walls at $y=\pm h$ (see figure~\ref{fig:q}).
The parameter $h$ is the width of the near-wall region that we 
would like to emphasize and the small parameter $\epsilon$ determines how quickly the function 
$q_{\epsilon,h}$ decays to zero far from the walls.

The POD modes are computed through the equations~\eqref{eq:pod} and~\eqref{eq:pod_eig}
except that in computing the covariance~\eqref{eq:pod2} instead of the terms $\vc u(\vc x,t)-\overline{\vc u}(\vc x)$, we use the mean-zero weighted velocity $\vc u_{\epsilon,h}(\vc x,t)$.
In the following, we set $\epsilon = 0.1$ to keep the decay smooth. We selected $h = 25\delta/110$ to emphasize the 
near-wall flow up to ignore any flow above $y^+\approx50$, where the minimal flow unit does 
not capture all length scales. 
This modified POD not only emphasizes the near-wall contributions, but also speeds up the convergence of the 
numerical optimization of problem~\eqref{eq:optim_pod}.

The POD modes were computed using 1000 snapshots taken from flows computed from 10 different randomly chosen, initial conditions. The snapshots were taken at intervals of 50 eddy turnover times to minimize any correlation in time. Fig. \ref{f:pod_modes_energy} shows the energy and cumulative energy of the POD modes. Almost 80\% of the energy is captured by the first six modes, indicating that near-wall dynamics emphasized by our weight function are low dimensional. 

\begin{figure}
\centering
\subfigure[Mode 1]{\includegraphics[width=.3\textwidth]{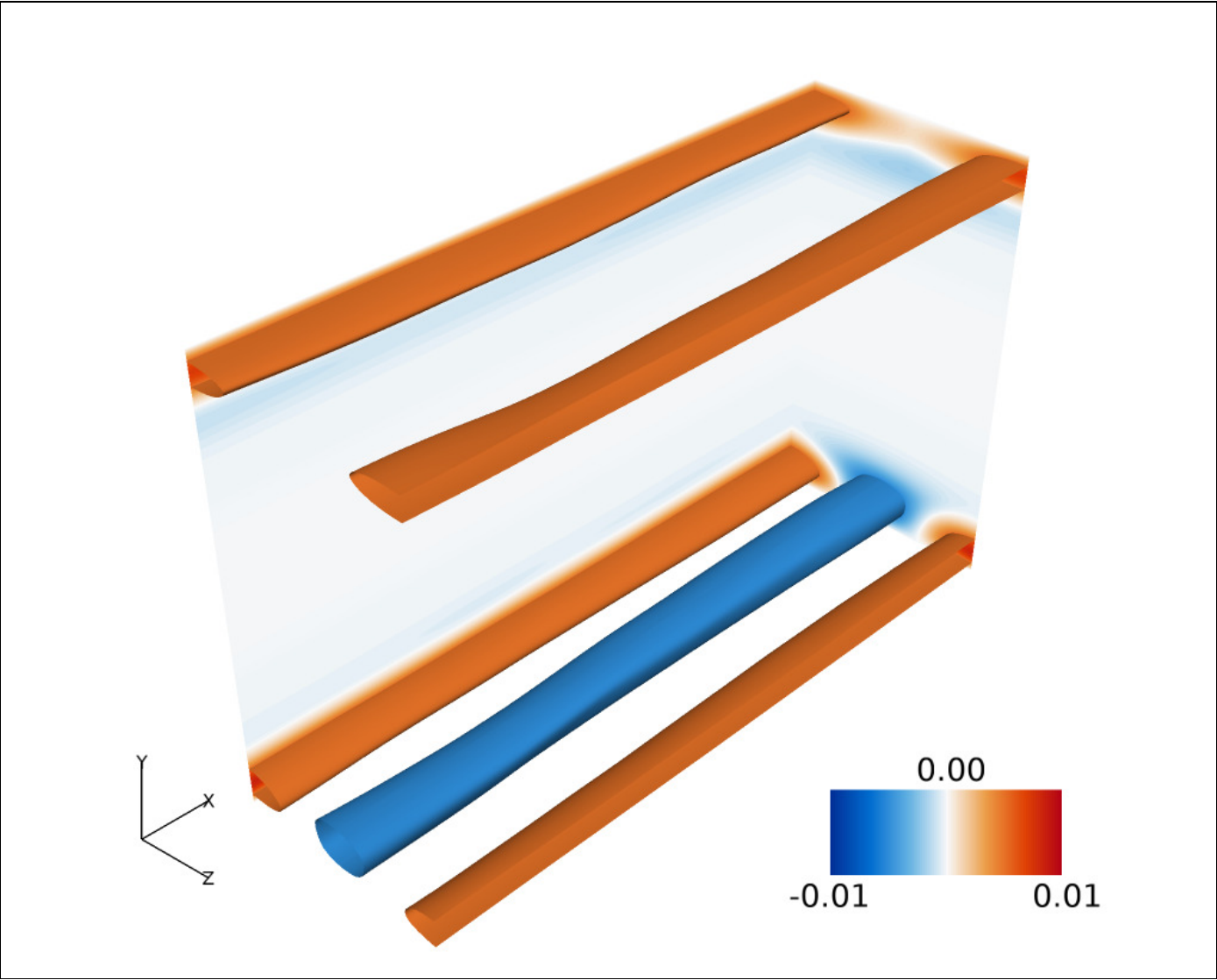}}
\subfigure[Mode 3]{\includegraphics[width=.3\textwidth]{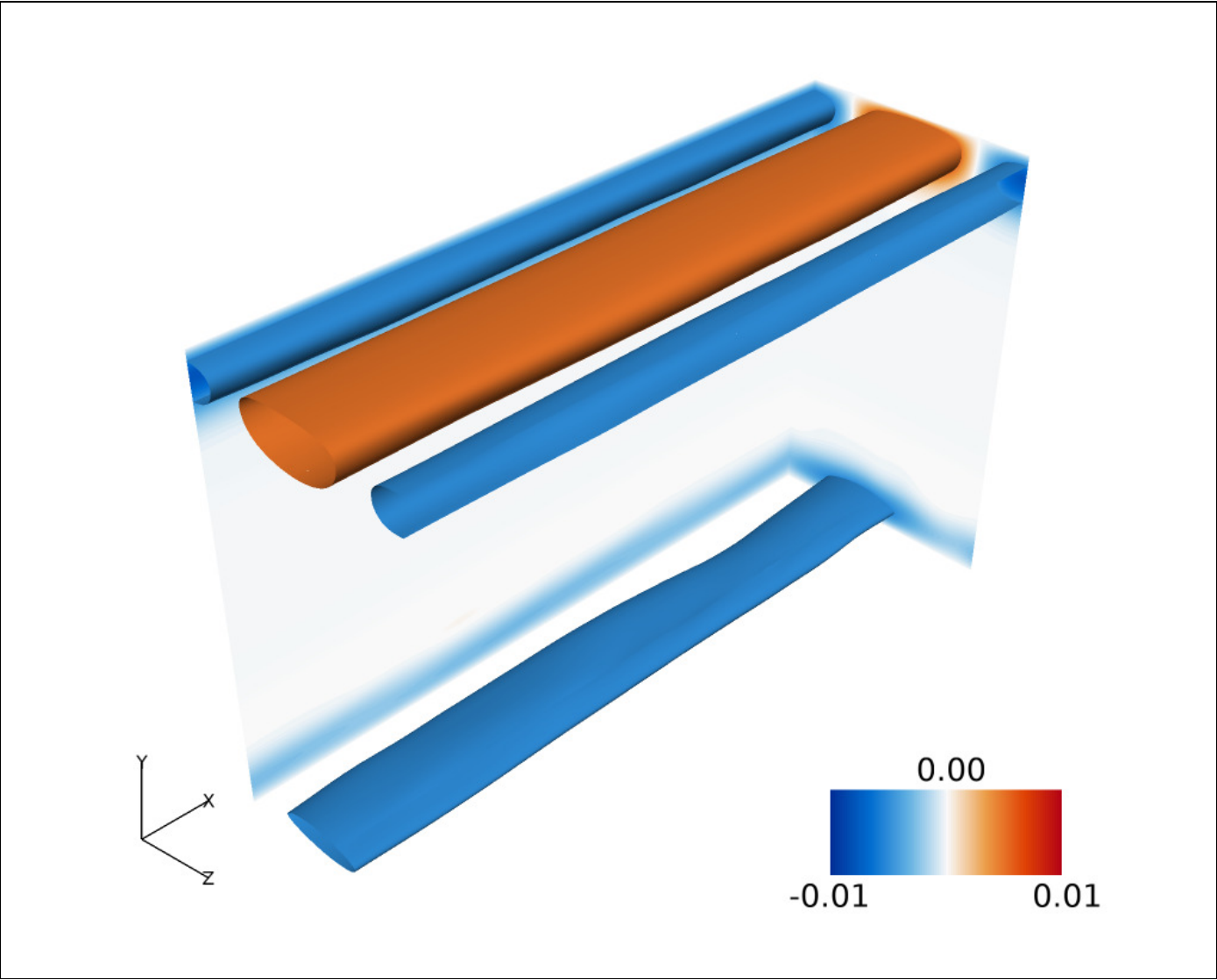}}
\subfigure[Mode 11]{\includegraphics[width=.3\textwidth]{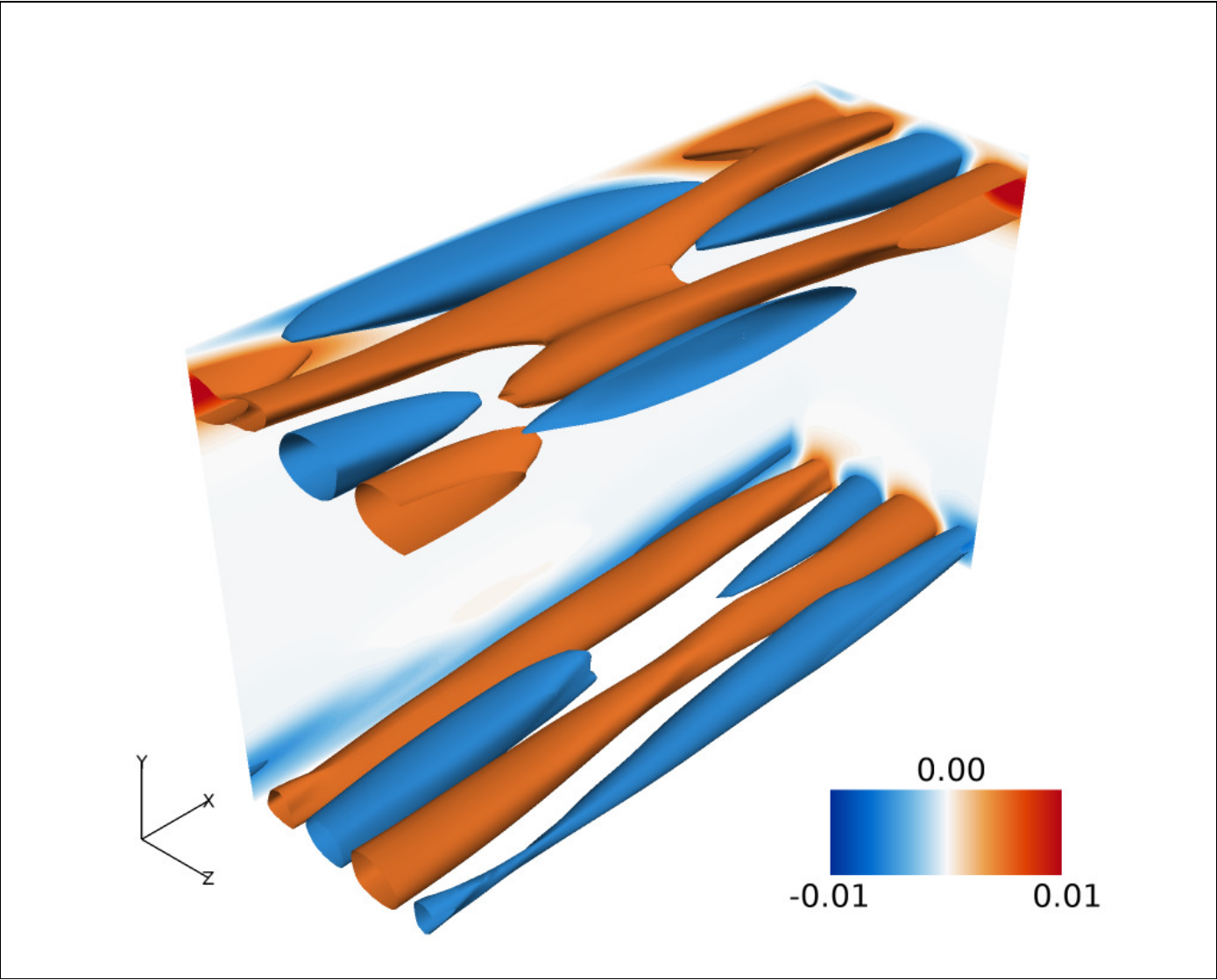}}
\caption{Contours and isosurfaces of axial velocity $u$ for several POD modes. Isosurfaces are defined at $u =\pm 0.01$. Mode indicies are defined as in Fig. \ref{f:pod_modes_energy}.}
\label{f:pod_modes_xmom}
\end{figure}

Fig. \ref{f:pod_modes_xmom} shows a few of these modes. The shape of the most energetic modes is dominated by long axial streaks which are know to be a main feature of the near-wall region \cite{Jimenez_2012}. The more energetic modes such as modes 1 and 3 contain wider streaks, while less energetic modes such as mode 11 contain thinner, less coherent streaks that meander slightly. 

Finally, we point out that an energy maximizing optimization similar to~\eqref{eq:optim} was used by Pringle and Kerswell~\cite{pringle2010} in 
the context of transition to turbulence in the pipe flow. The optimal perturbations to the laminar state that
lead to turbulence were found as solutions to an optimization problem. 
In that context, the perturbations to a known state (i.e., the laminar state) 
are sought and therefore the knowledge of the turbulent attractor 
does not enter the optimization problem. In contrast, our study demands that the initial states belong to 
the turbulent attractor as enforced through the constraint~\eqref{eq:const_2} and implemented numerically in this section.
Similarly, Farano et al.~\cite{farano2017} investigated the observed bursts in a turbulent flow similar to our channel flow. However, they only
require that the energy of the optimal state is prescribed, i.e., $E(\vc u_0)=E_0$ for a prescribed energy level $E_0$.
This does not necessarily imply that the optimal state belongs to the attractor and therefore unphysical 
optimizers are not ruled out.

\subsection{Numerical implementation}

The optimization problem, equation~\eqref{eq:optim_pod}, was solved using the Python package scipy \cite{Scipy}. Specifically, the ``optimize'' package was used and the optimization was carried out using Sequential Least SQuares Programming (SLSQP). The first constraint,~\eqref{eq:const1_pod}, is implicitly satisfied by the flow solver, which is called from Python using the ``multiprocessing'' module. The second and third constraints~\eqref{eq:const2_pod}--\eqref{eq:const3_pod} and their gradients are implemented directly in Python. The convergence tolerance for the objective function was set to $10^{-7}$. Otherwise the default convergence criterion were used. An adjoint solver was used to minimize the cost of computing the gradient of the objective function \eqref{eq:max_pod}. The DGSEM solver has a dual consistent, discrete adjoint formulation, details of which are discussed in Refs. \cite{Ceze_Diosady_Murman_2016,Carton_Diosady_2018}. This adjoint solver allows us to compute the gradient at a computational cost similar to solving the primal, which is much cheaper than using finite differences to compute the gradient with respect to all $n=50$ POD modes. 

\subsection{Optimization results}

The optimization was run from five different initial guesses $\vc u_0$ and with three different integration times of $\tau=52.5t_e$, $\tau=105t_e$, and $\tau=210t_e$. The integration times correspond to roughly $1/8$, $1/4$, and $1/2$ of the time scale of laminarization event similar to the one shown in Fig. \ref{f:ee_ke_diss}. All five optimizations with $\tau=52.5t_e$, and the optimization with $\tau=105t_e$ computed very similar optimal solutions $\vc u_0^*$. The optimization with an integration time of $\tau=210t_e$ failed to converge because the adjoint grew very large in magnitude, and the gradient caused the optimizer to consider a non-physical solution in the ensuing line search.

\begin{figure}
\centering
\includegraphics[width=.8\textwidth]{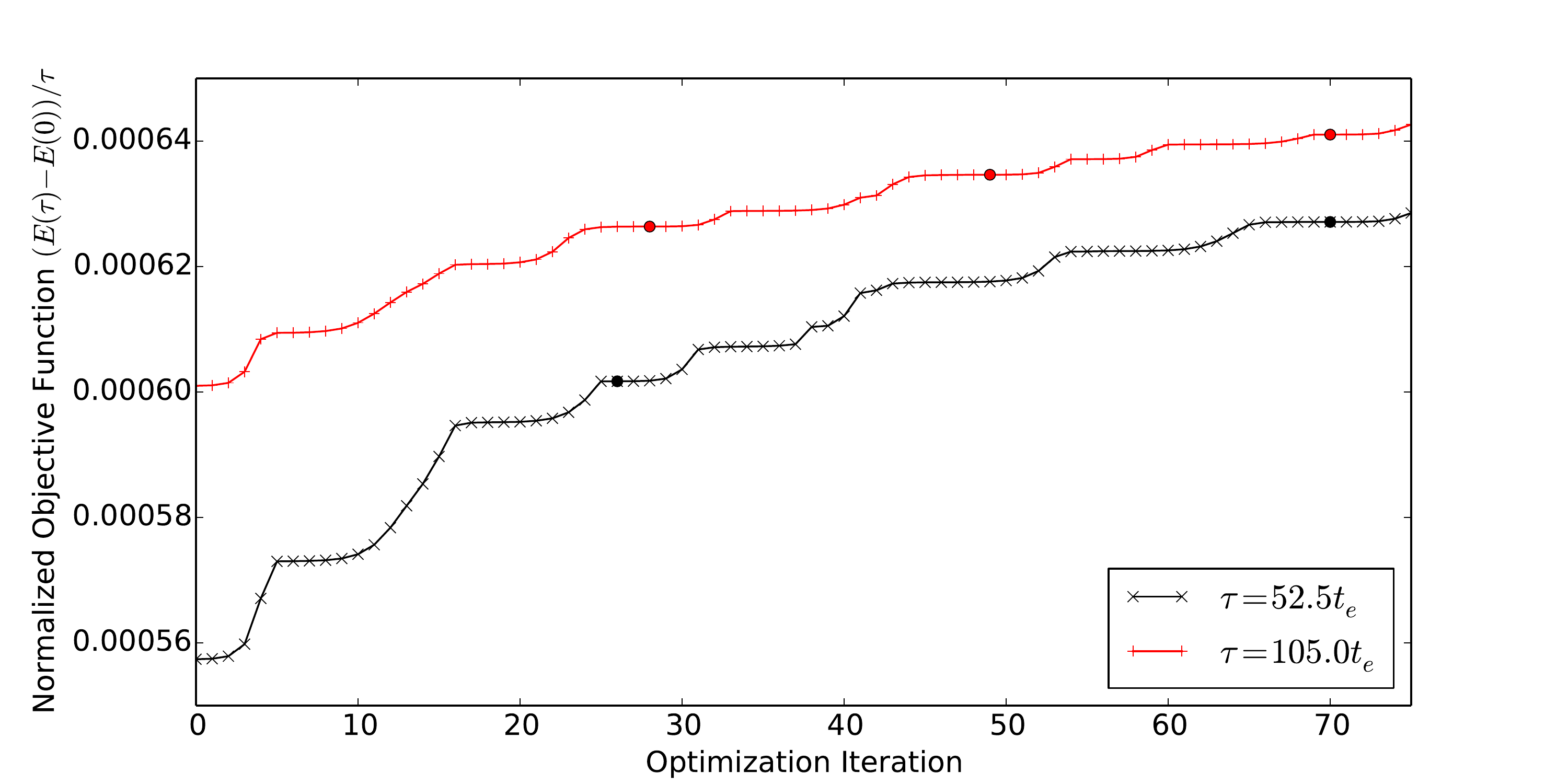}
\caption{Convergence of the objective function for the same initial guess but two different choices of $\tau$. Solid dots represent instances when the optimizer was restarted. }
\label{f:objective_convergence}
\end{figure}

\begin{figure}
\centering
\includegraphics[width=.9\textwidth]{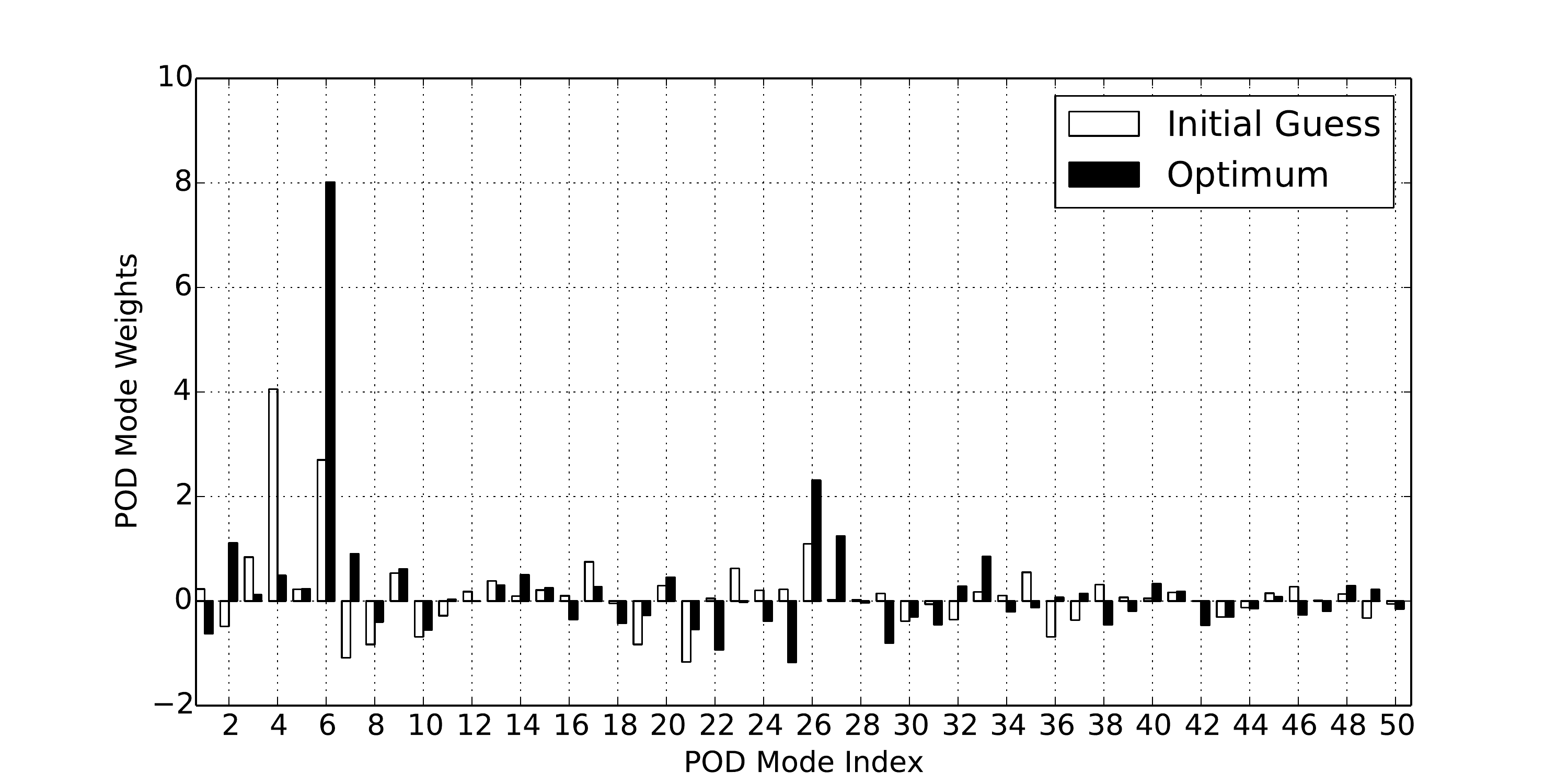}
\caption{POD mode weights for the initial guess and optimal solution.}
\label{f:optimal_pod_weights}
\end{figure}

\begin{figure}
\centering
\subfigure[Mode 6]{\includegraphics[width=.3\textwidth]{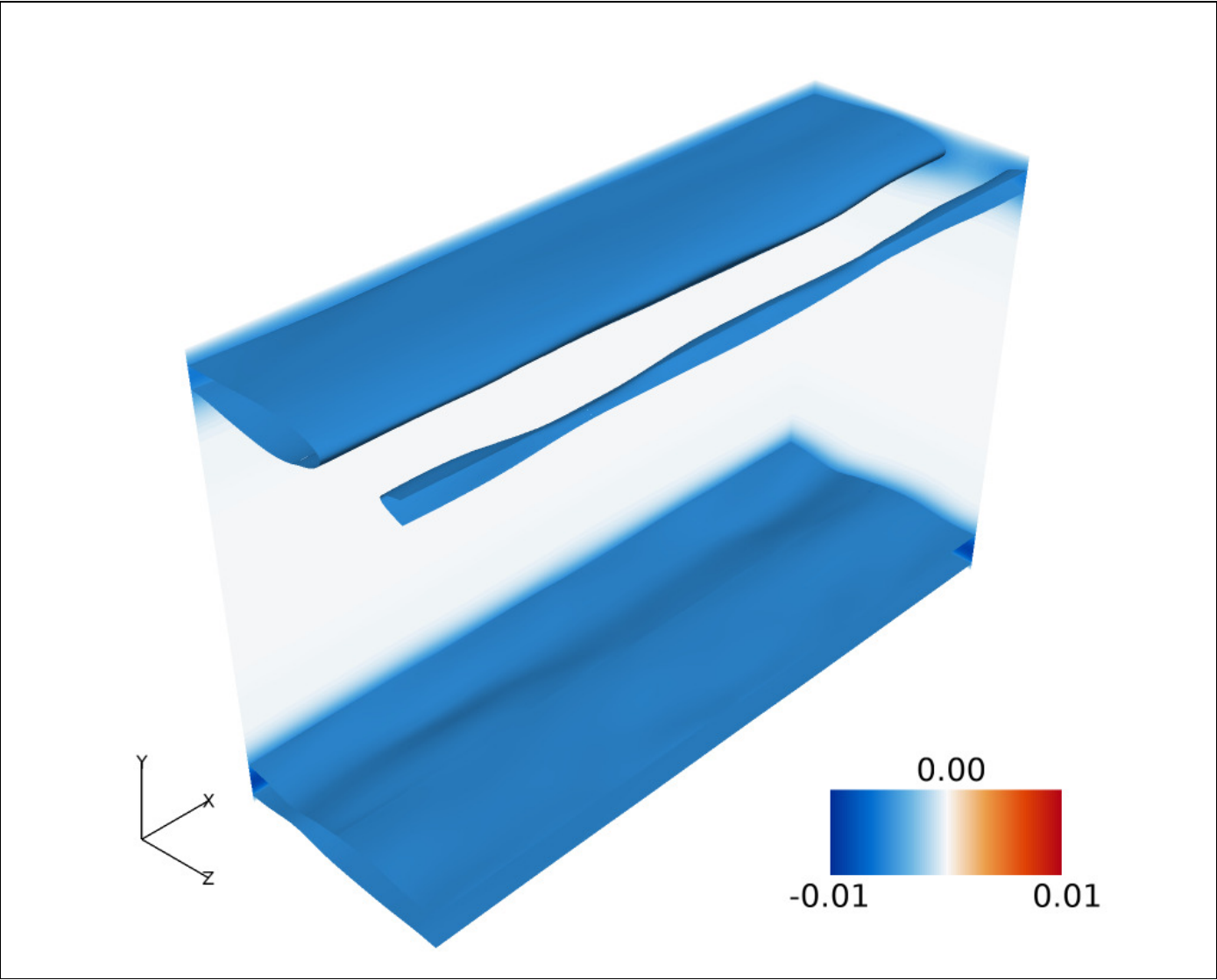}}
\subfigure[Mode 6 Profile]{\includegraphics[width=.3\textwidth]{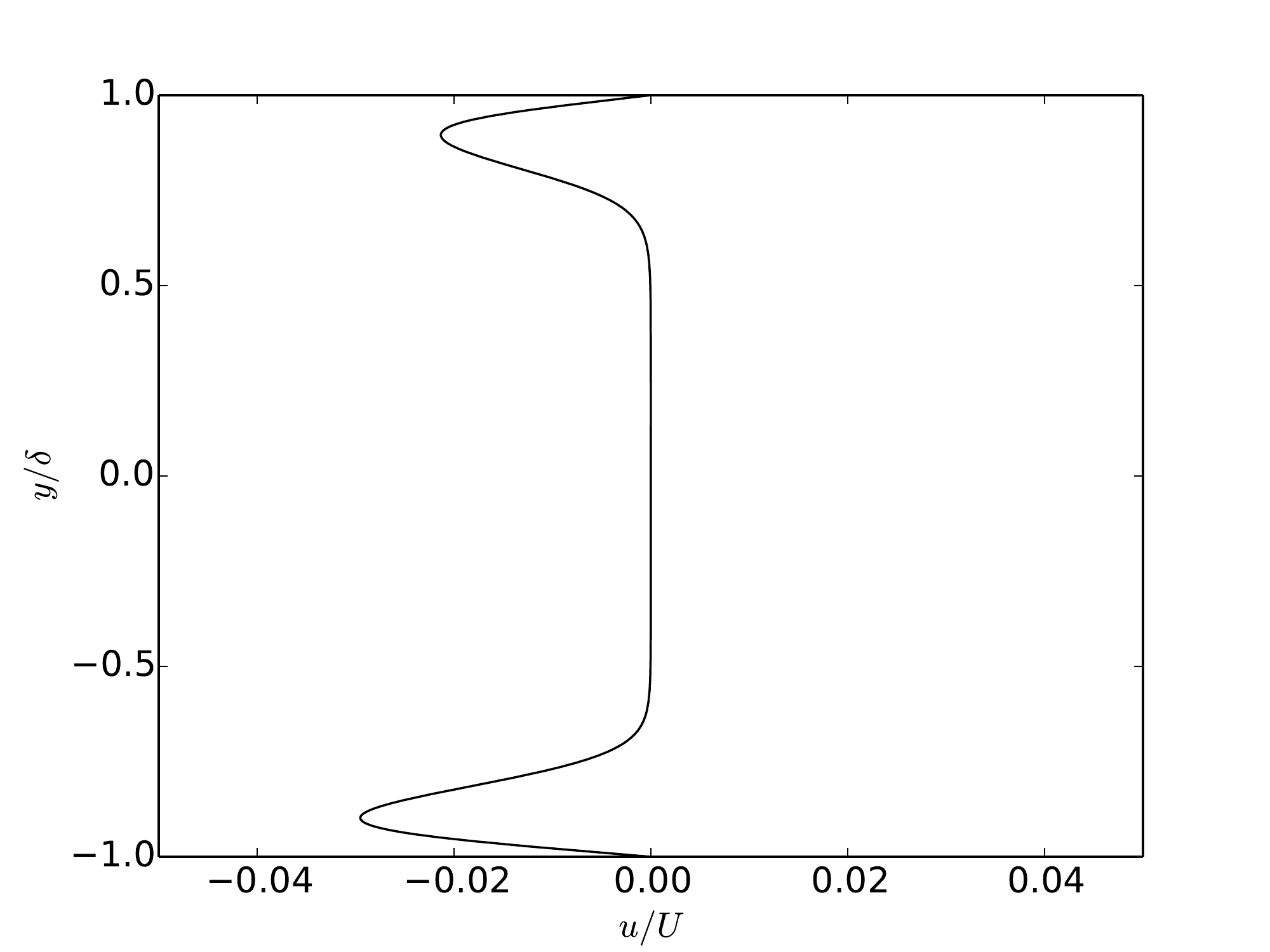}} \\
\subfigure[Mode 26]{\includegraphics[width=.3\textwidth]{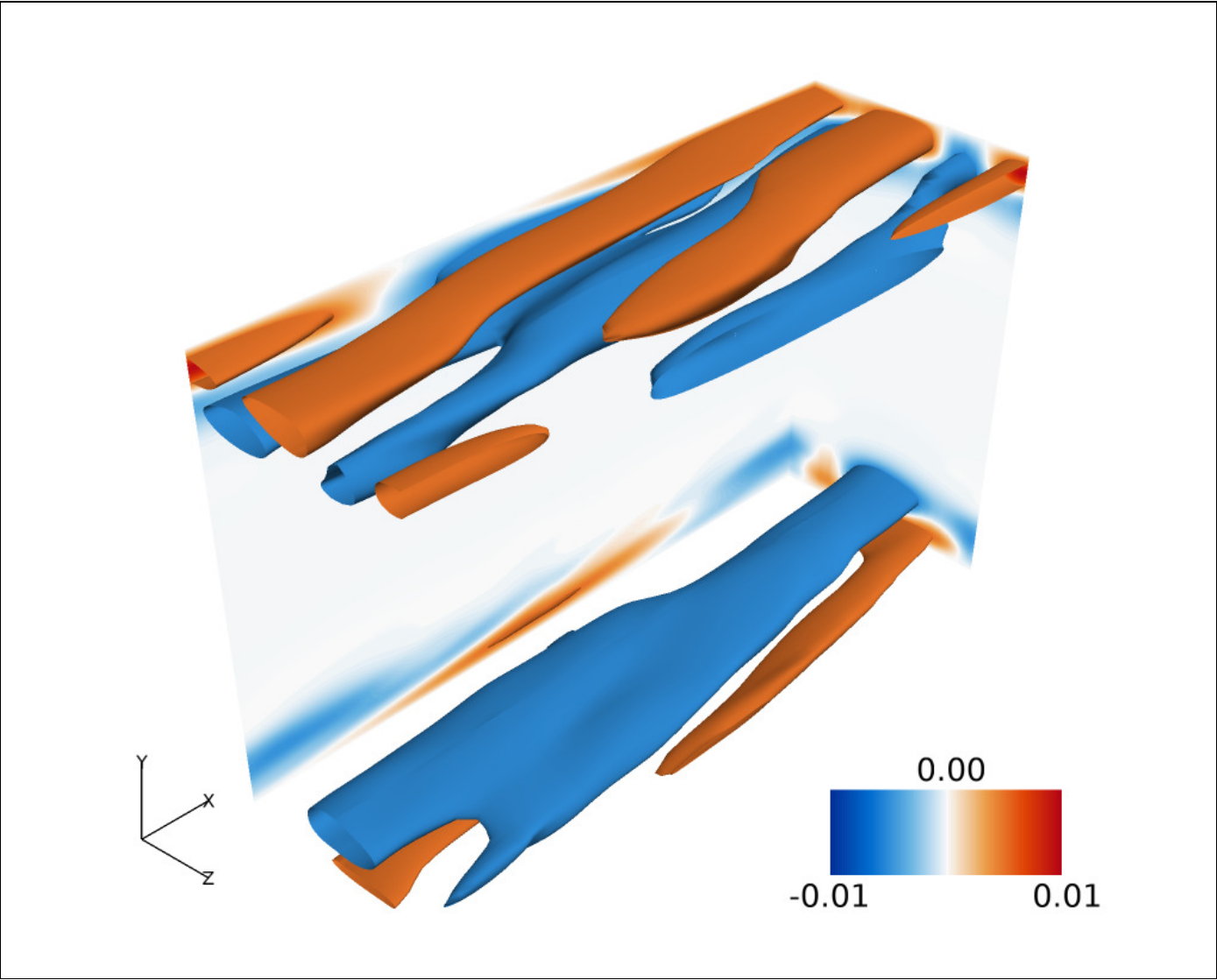}}
\subfigure[Mode 26 Profile]{\includegraphics[width=.3\textwidth]{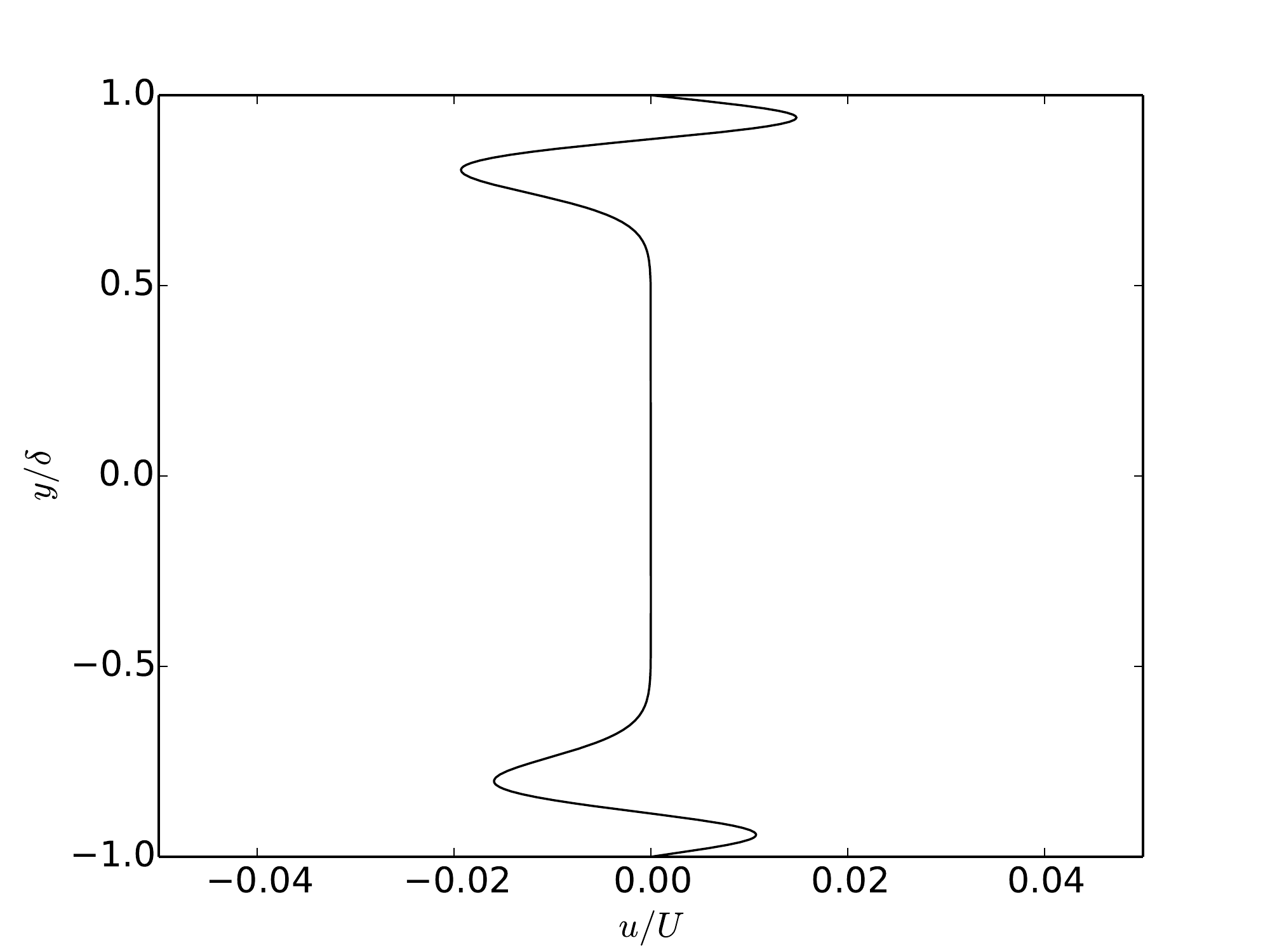}}
\caption{Contours, isosurfaces, and spatially averaged profiles of axial velocity for the two most energetic POD modes in the optimal solution to Eq. \eqref{eq:optim_pod}. Isosurfaces are defined at $u \pm 0.01$. }
\label{f:optimal_solution_modes}
\end{figure}

\begin{figure}
\centering
\subfigure[]{\includegraphics[width=.6\textwidth]{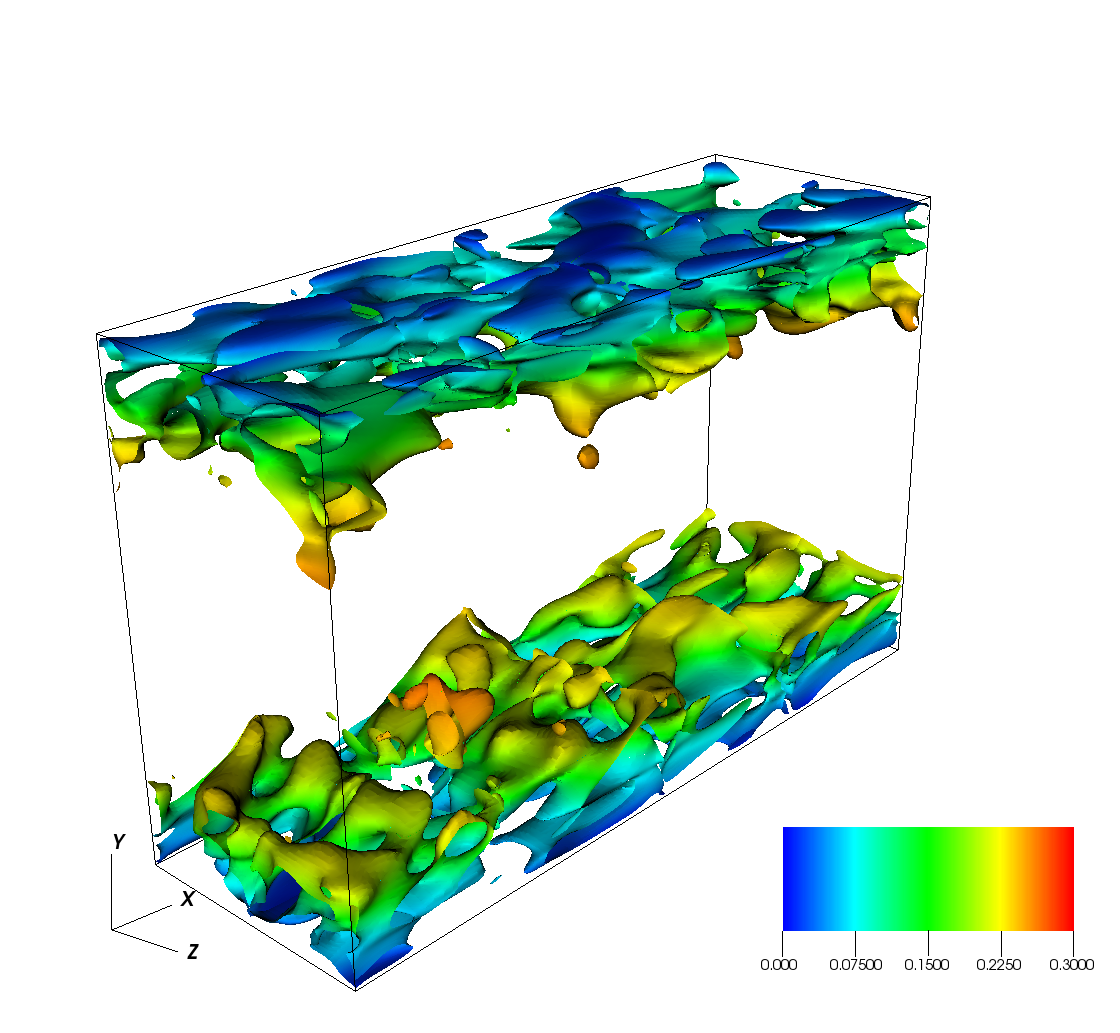}}
\subfigure[]{\includegraphics[width=.4\textwidth]{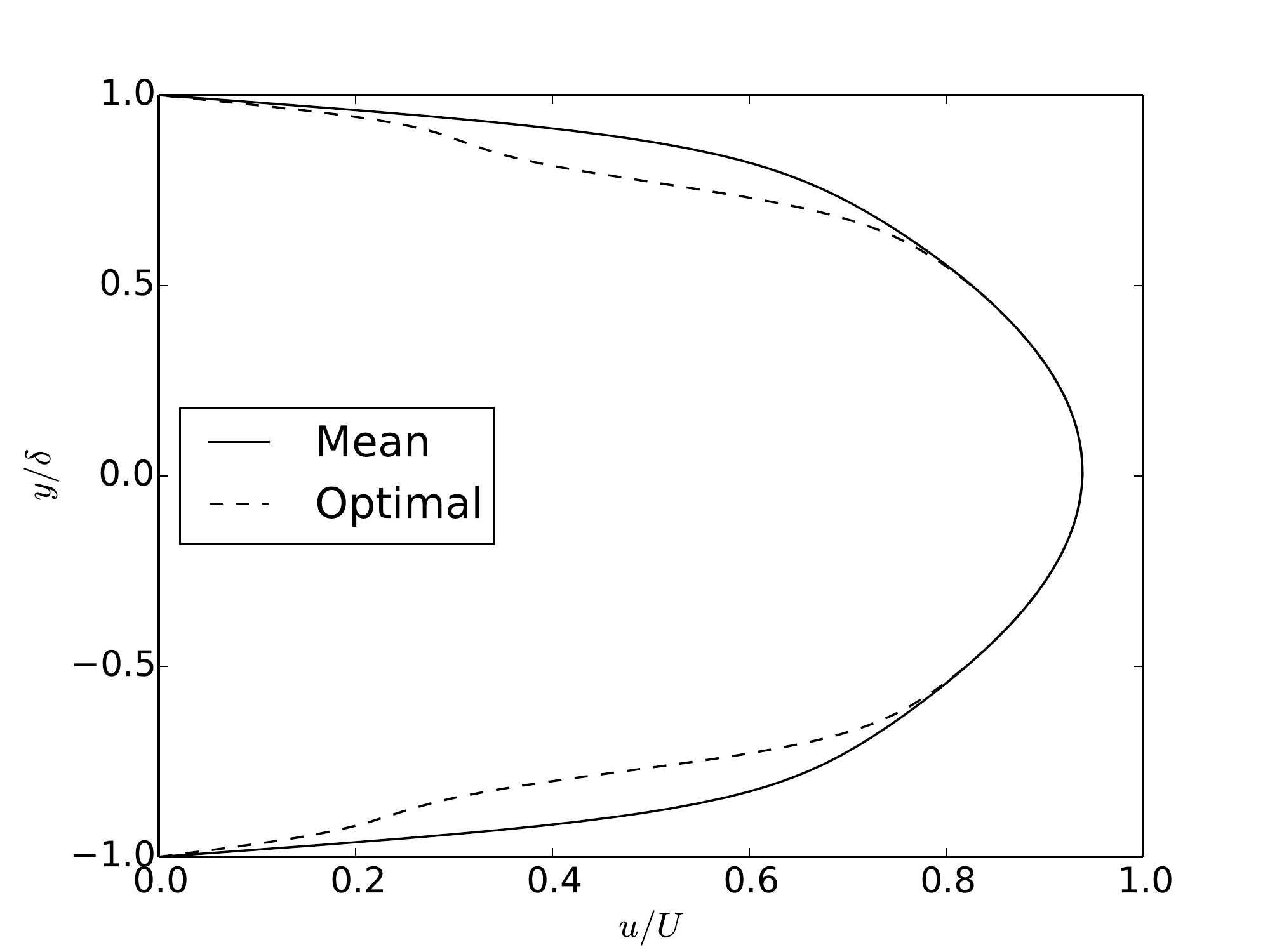}}
\subfigure[]{\includegraphics[width=.4\textwidth]{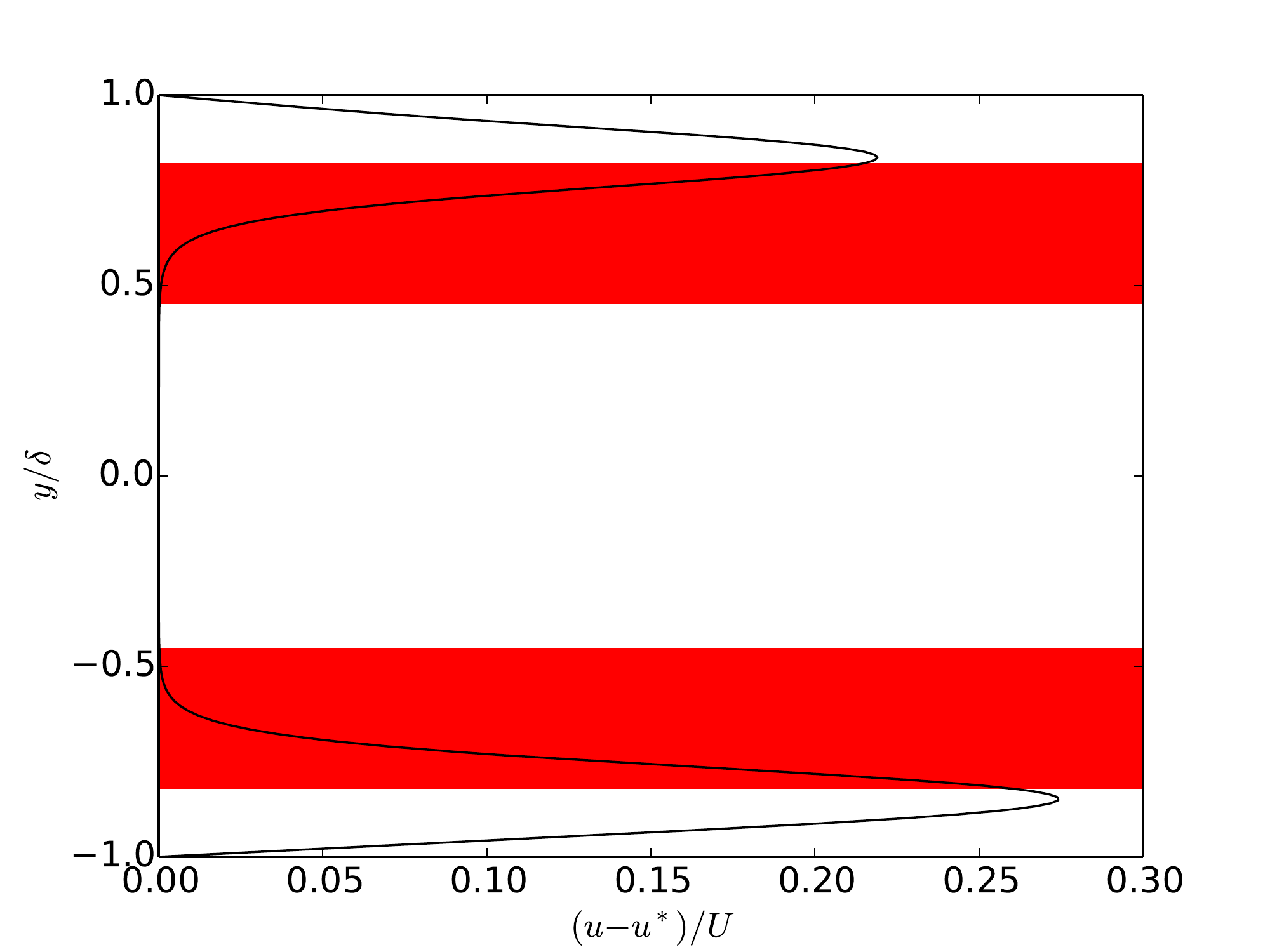}}
\caption{Optimal solution $\vc u_0^*$ to equation \eqref{eq:optim_pod}. 
(a) Q criterion isosurface for $Q=0.0005$, colored by the streamwise velocity
(b) Spatially averaged streamwise velocity profile with the mean axial velocity profile.
(c) Spatially averaged streamwise velocity profile difference. Red shaded boxes indicate regions from $y^+=20$ to $y^+=60$ units away from the wall.}
\label{f:optimal_solution}
\end{figure}

Fig. \ref{f:objective_convergence} shows the convergence history of a typical optimization for $\tau=52.5t_e$. The corresponding POD mode weights for the initial guess and optimal solution are shown in  Fig. \ref{f:optimal_pod_weights}. The two largest POD modes in the optimal solution are modes 6 and 26, shown in Fig. \ref{f:optimal_solution_modes}. Together, these modes create the axial velocity deficit in the near wall region shown in Fig. \ref{f:optimal_solution}. The maximum velocity deficit occurs roughly $y^+=18$ units from each wall, and decays to zero roughly $y^+ \approx 55$ units from the wall. 

The region in which the velocity deficit occurs is known to have a major impact on near-wall turbulence. In \cite{Jimenez_Pinelli_1999} it was observed that damping axial velocity streaks or quasi-streamwise vortex structures between $y^+ \approx 20$ to $y^+ \approx 60$ led to laminarization of the near wall region. It appears that our optimal solution modifies the lower portion of this range. The relative uniformity of the axial velocity deficit results in the absence of axial velocity streaks and quasi-streamwise vortex structures. This is in contrast with the flow snapshots in Fig. \ref{f:extreme_event_snapshots}, where the presence of low- and high- axial velocity streaks can be inferred from streaks in the wall shear stress, and quasi-axial structures are revealed by the q-criterion iso-contours. Without the axial velocity streaks and quasi-streamwise vorticity, the ``streak-cycle'' mechanism for the regeneration of near-wall turbulent fluctuations is broken and the flow laminarizes. 

Therefore, although the optimal solution $\vc u_0^*$ is turbulent, the absence of axial velocity streaks and quasi-streamwise vortices make it a precursor for a flow laminarization. By tracking how close a given state is to $\vc u_0^*$, we can determine if a laminarization event is likely to occur or not.

\section{Predicting extreme events}\label{sec:pred}
\begin{figure}
\centering
\subfigure[]{\includegraphics[width=.49\textwidth]{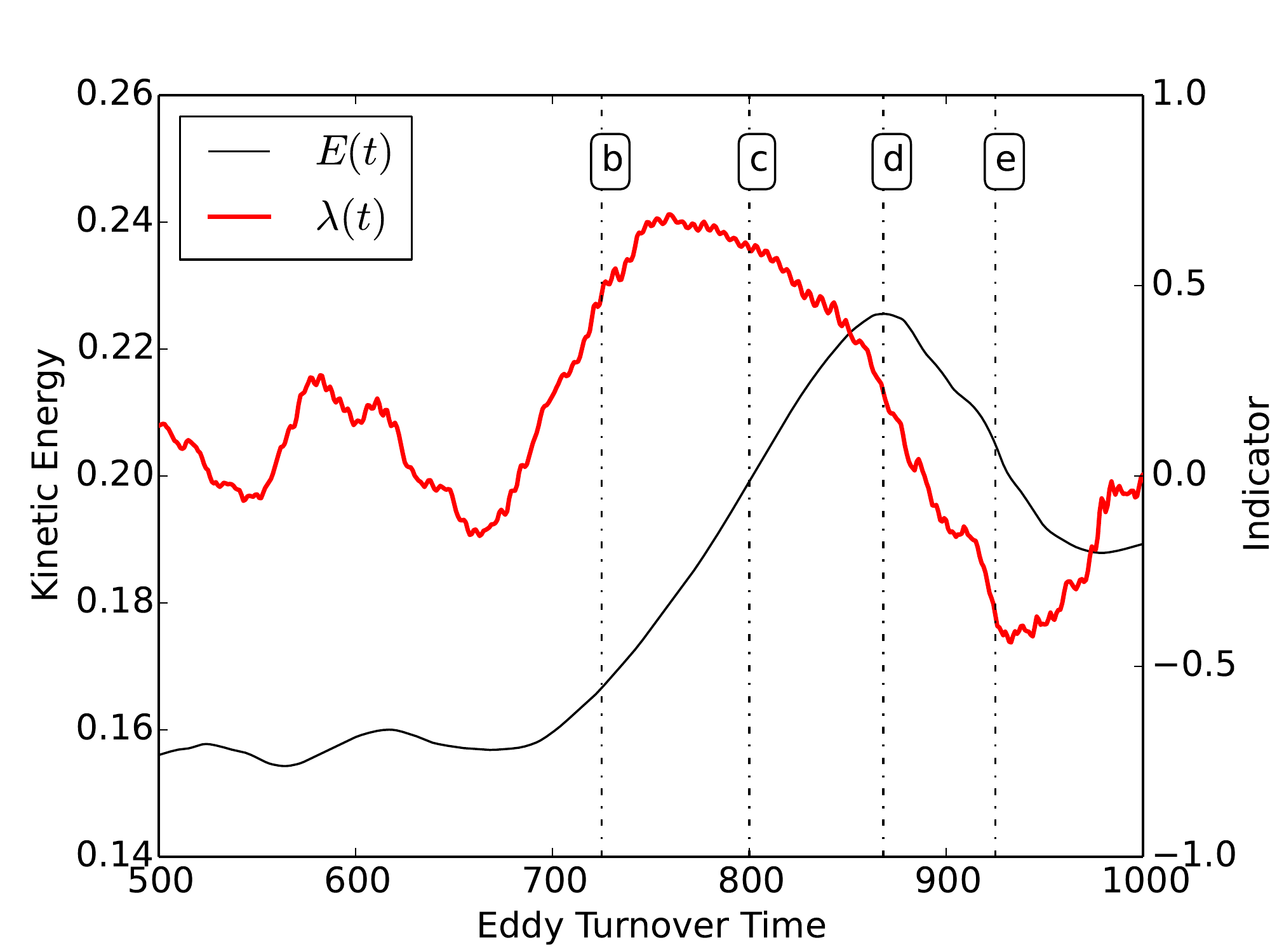}}
\subfigure[]{\includegraphics[width=.49\textwidth]{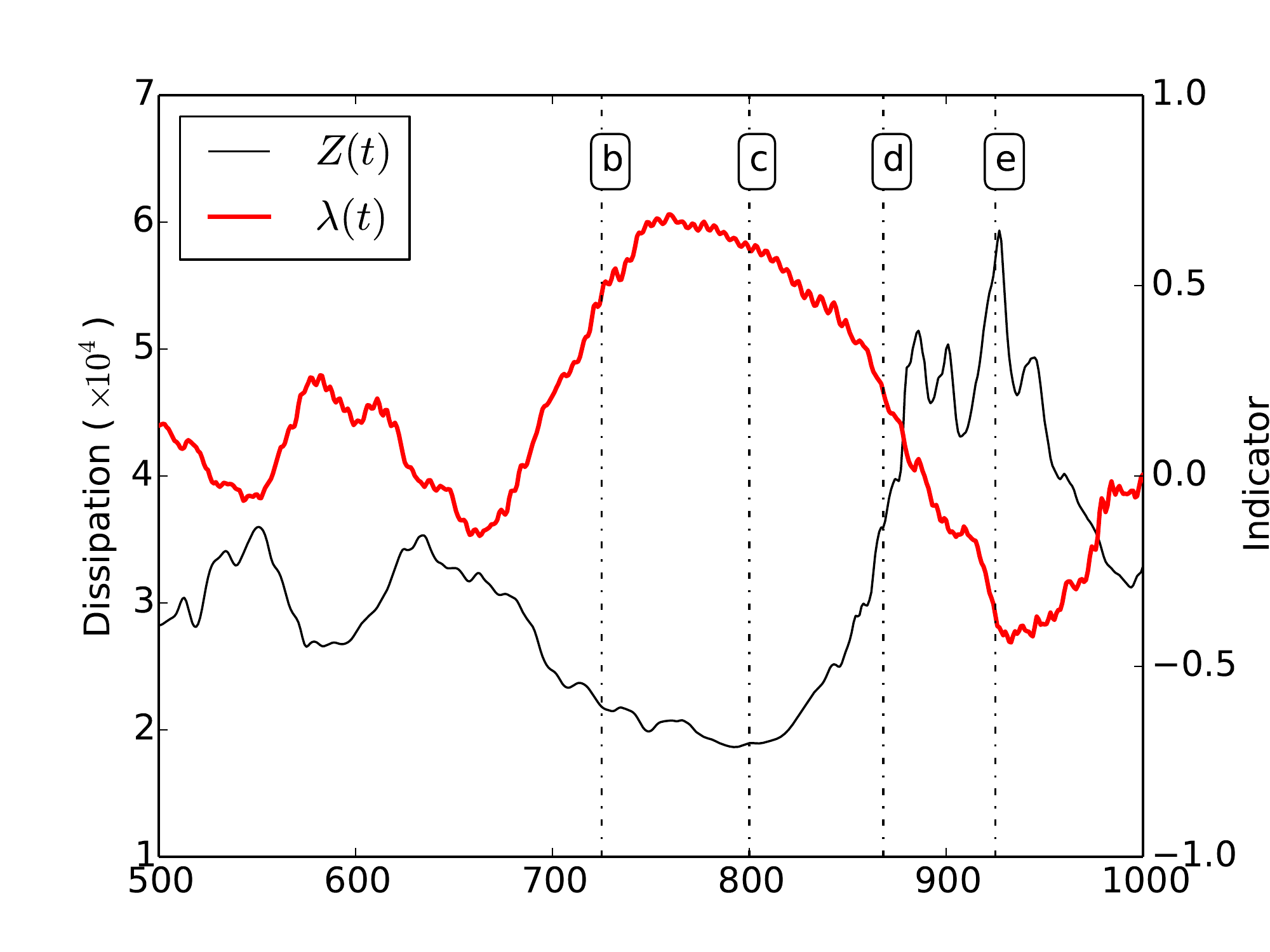}}
\caption{Time evolution of  the indicator $\lambda(t)$ with $E(t)$ and $Z(t)$  over the same time horizon as shown in Fig. \ref{f:ee_ke_diss}.   The vertical lines indicate the times that snapshots in Fig. \ref{f:extreme_event_snapshots} correspond to. The first and last snapshot correspond to the start and end of the time horizon shown above.  }
\label{f:ee_ke_diss_lam}
\end{figure}

Figure~\ref{f:ee_ke_diss_lam} shows a close-up of the evolution of 
energy $E$ and dissipation $Z$ together with the indicator 
\begin{equation}\label{eq:indicator}
\lambda =\frac{ \left\langle \vc u - \overline{\vc u} ,\vc u_0^* - \overline{\vc u}  \right\rangle}{\| \vc u - \overline{\vc u} \|_2 \| \vc u_0^* - \overline{\vc u}  \|_2},
\end{equation}
at $Re=2200$.
Large values of energy and dissipation are proceeded with relatively large values of the indicator. 
This behavior turns out to be generic and not specific to this time window. As a result, 
one can use the indicator $\lambda$ to predict the upcoming extreme events in the channel flow.
In order to quantify these predictions, we first review some statistical tools in Section~\ref{sec:stat_prelim}. 
Subsequently, in Section~\ref{sec:pred_results}, we apply these tools to longterm simulations of the 
channel flow and report the results. 

\subsection{Statistical preliminaries}\label{sec:stat_prelim}
In this section, we show that the trigger state obtained previously can be used for the prediction of 
the extreme events in the channel flow. In order to make quantitative statements, we 
use joint and conditional statistics between the trigger mode and the energetic observables of the
turbulent flow, namely kinetic energy and energy dissipation rate.

For a given random variable $X_t$, we would like to find an indicator $Y_t$ (another random variable)
whose values signal an upcoming extreme event of $X_t$.
We identify extreme events of $X_t$ as any instant where $X_t>x_e$ for a 
prescribed extreme value threshold $x_e$.
First we define the maximum value of the random 
variable $X_t$ over a future time interval $[t+t_0,t+t_0+\Delta t]$
for some $t_0,\Delta t\ge 0$,
\begin{equation}
\tilde X_t(t_0,\Delta t) = \max_{s\in [t+t_0,t+t_0+\Delta t]} X_s
\label{eq:rv_max}
\end{equation}
The maximum $\tilde X_t(t_0,\Delta t)$ is a new random variable which depends on the
parameters $t_0$ and $\Delta t$. At any time $t$, 
$\tilde X_t(t_0,\Delta t)$ measures the maximum value that $X_t$ will take over 
the future time interval $[t+t_0,t+t_0+\Delta t]$.
For notational simplicity, we omit the parameters
$t_0$ and $\Delta t$ and simply write $\tilde X_t$. 

The joint probability distribution of the pair $(\tilde X_t,Y_{t})$ is defined by
\begin{align}\label{eq:jointPDF}
F_{\tilde X_t,Y_{t}}(x,y)&=\mathbb P(\tilde X_t\leq x,Y_{t}\leq y)\nonumber\\
&=\int_{-\infty}^{x}\int_{-\infty}^{y}p_{\tilde X_t,Y_{t}}(x,y)\id x\id y,
\end{align}
where $p_{\tilde X_t,Y_{t}}$ is the probability density associated with the probability distribution $F_{\tilde X_t,Y_{t}}$. The probability density $p_{\tilde X_t,Y_t}(x,y)$
measures the probability that at time $t$ we observe $Y_t=y$ and $\tilde X_t=x$.

The conditional probability of $\tilde X_t=x$ given that $Y_t=y$ is defined through the Bayes' formula
\begin{equation}\label{eq:condPDF}
p_{\tilde X_t|Y_t} = \frac{p_{\tilde X_t,Y_t}}{p_{Y_t}},
\end{equation}
where $p_{Y_t}$ is the probability density associated with the random variable $Y_t$.
We use the conditional PDF $p_{\tilde X_t|Y_t}$ to quantify the extent to which the
behavior of $Y_t$ is indicative of the extreme events of $X_t$ over the 
future time interval $[t+t_0,t+t_0+\Delta t]$.
More precisely, given an extreme event threshold $x_e$, we define the probability of upcoming extreme events by
\begin{equation}\label{eq:Pee}
P_{ee}(y) = \int_{x_e}^\infty p_{\tilde X_t|Y_t}(x,y)\id x.
\end{equation}
This quantity measures the probability of an extreme event over the future time interval 
$[t+t_0,t+t_0+\Delta t]$ given the current value of the indicator $Y_t=y$.

For a reliable indicator, $P_{ee}$ should be monotonic so that the
probability of upcoming extreme events increases with $y$. 
More precisely, $P_{ee}$ should be nearly zero for small values of $y$
and increase monotonically towards 1 as $y$ increases. 
We predict an upcoming extreme only if $P_{ee}>0.5$. 
This defines an extreme event threshold $y_e$ for the indicator $Y_t$
where $P_{ee}(y_e)=0.5$. 
If $Y_t>y_e$ an upcoming extreme event is predicted and conversely if $Y_t<y_e$
it is predicted that no extreme events will occur over the future time interval $[t+t_0,t+t_0+\Delta t]$.

This classification leads to four possible prediction outcomes in terms of the indicator value $Y_t$ and
the future observable values $\tilde X_t$:
\begin{align}
\mbox{Correct Rejection (CR):} &\qquad \tilde X_t<x_e\quad\mbox{given}\quad  Y_t<y_e,\nonumber\\
\mbox{Correct Prediction (CP):} &\qquad \tilde X_t>x_e\quad\mbox{given}\quad Y_t>y_e,\nonumber\\
\mbox{False Negatives (FN):} &\qquad \tilde X_t>x_e\quad\mbox{given}\quad Y_t<y_e,\nonumber\\
\mbox{False Positive (FP):} &\qquad \tilde X_t<x_e\quad\mbox{given}\quad Y_t>y_e.
\end{align}
Therefore, the skill of an indicator for predicting upcoming extreme events can be quantifies as follows
\begin{subequations}\label{eq:rs}
\begin{align}
\mbox{Rate of successful predictions} &= \frac{\mbox{CP}}{\mbox{CP}+\mbox{FN}}\nonumber\\
& =\frac{\int_{x_e}^\infty\int_{y_e}^\infty p_{\tilde X_t|Y_t}(x,y)\id y\id x}{\int_{x_e}^\infty\int_{-\infty}^\infty p_{\tilde X_t|Y_t}(x,y)\id y\id x}
\end{align}
\begin{align}
\mbox{Rate of successful rejections} &= \frac{\mbox{CR}}{\mbox{CR}+\mbox{FP}}\nonumber\\
& =\frac{\int_{-\infty}^{x_e}\int_{-\infty}^{y_e} p_{\tilde X_t|Y_t}(x,y)\id y\id x}{\int_{-\infty}^{x_e}\int_{-\infty}^{\infty} p_{\tilde X_t|Y_t}(x,y)\id y\id x}
\end{align}
\end{subequations}

A skillful indicator is one that returns relatively small percentage of false negatives (respectively, false positives) 
compared to the number of correct predictions (respectively, correct rejections).
In the following, we use the statistical quantities introduced above to quantify the predictive skill of the 
indicator~\eqref{eq:indicator}.

\subsection{Prediction results}\label{sec:pred_results}
\begin{figure}
\centering
\subfigure[]{\includegraphics[width=.3\textwidth]{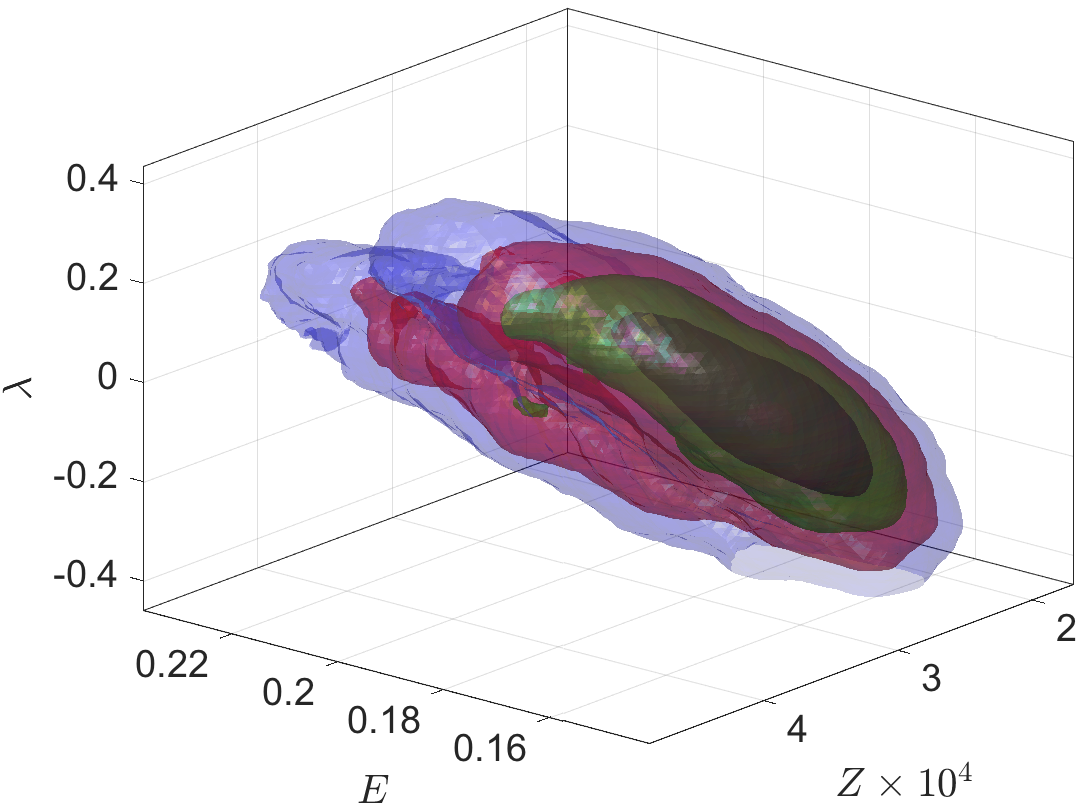}}
\subfigure[]{\includegraphics[width=.3\textwidth]{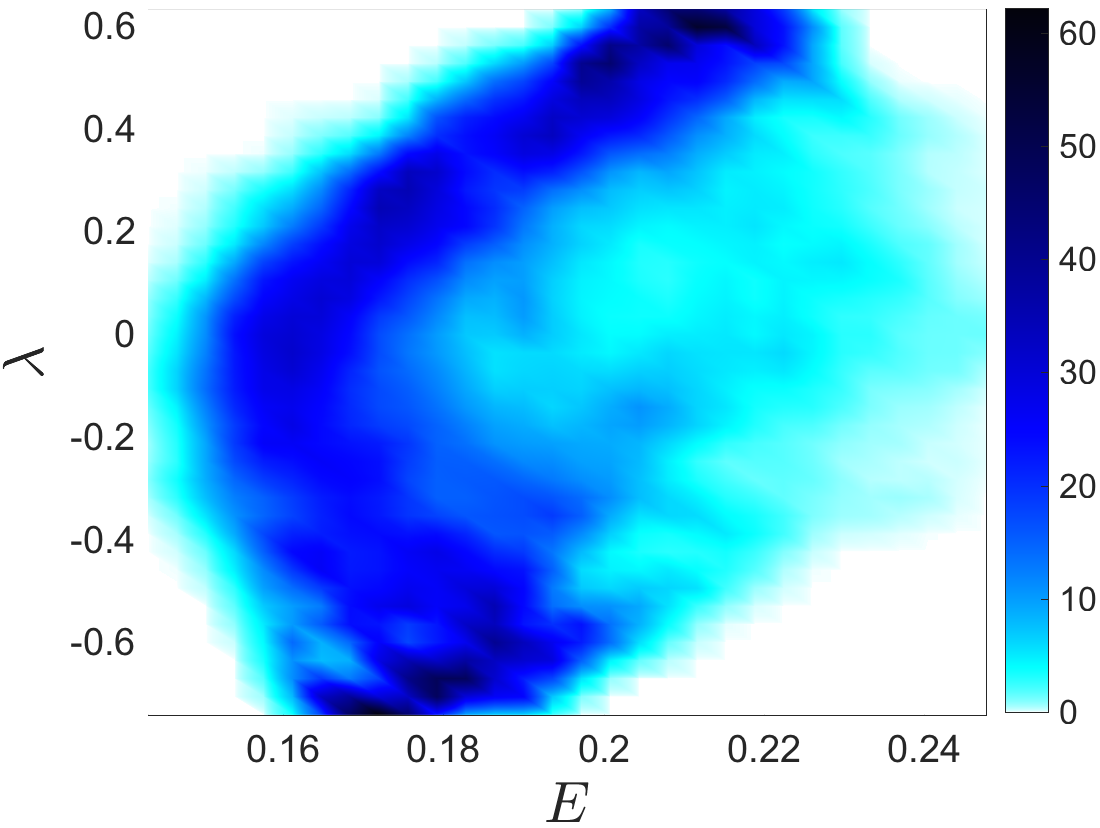}}
\subfigure[]{\includegraphics[width=.3\textwidth]{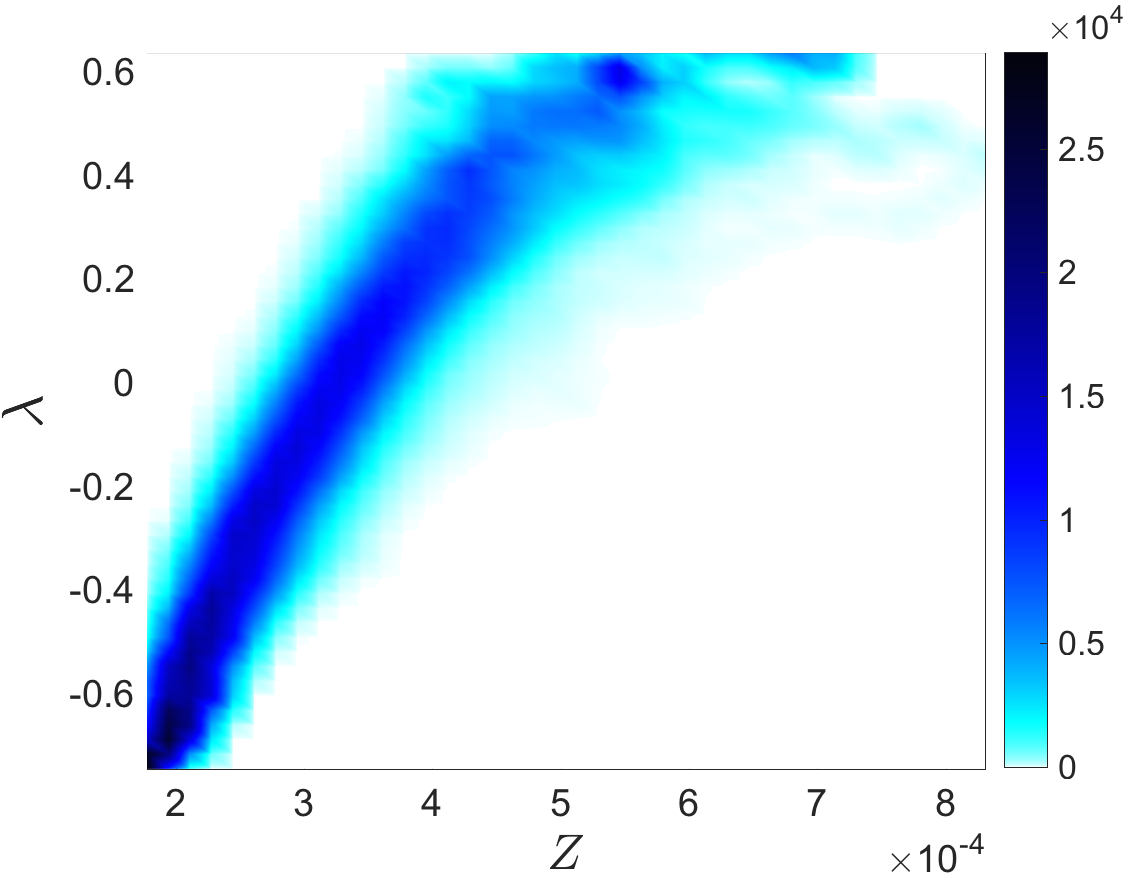}}
\caption{$Re=2200$. 
(a) Joint PDF of the kinetic energy $E$, energy dissipation $Z$ and the indicator $\lambda$.
(b) Conditional PDF $p_{E|\lambda}$. 
(c) Conditional PDF $p_{Z|\lambda}$.
}
\label{fig:R2200_dt0}
\end{figure}

We first present the joint and conditional statistics for the energy $E$, 
dissipation rate $Z$ and the indicator $\lambda$.
In this first step, we do not include any time shifts, thus setting
$t_0=\Delta t=0$ so that $\tilde X_t=X_t$ in equation~\eqref{eq:rv_max}. 
Figure~\ref{fig:R2200_dt0} shows the joint and conditional PDFs of the
indicator versus energy $E$ and the energy dissipation $Z$
at $Re=2200$. This figure is generated from an ensemble of longterm simulations with data recorded
every one eddy turnover time collecting a total of $106,063$ data points.

The shape of the conditional PDFs $p_{E|\lambda}$ and $p_{Z|\lambda}$
shows that the extreme values of the indicator $\lambda$ 
correlate strongly with relatively large energy and large dissipation episodes (see Fig.~\ref{fig:R2200_dt0}(a,b)).
We also note that that the correlation is much stronger for dissipation and
that $\lambda$ tends to increase monotonically with the dissipation $Z$. 

\begin{figure}
\centering
\includegraphics[width=.65\textwidth]{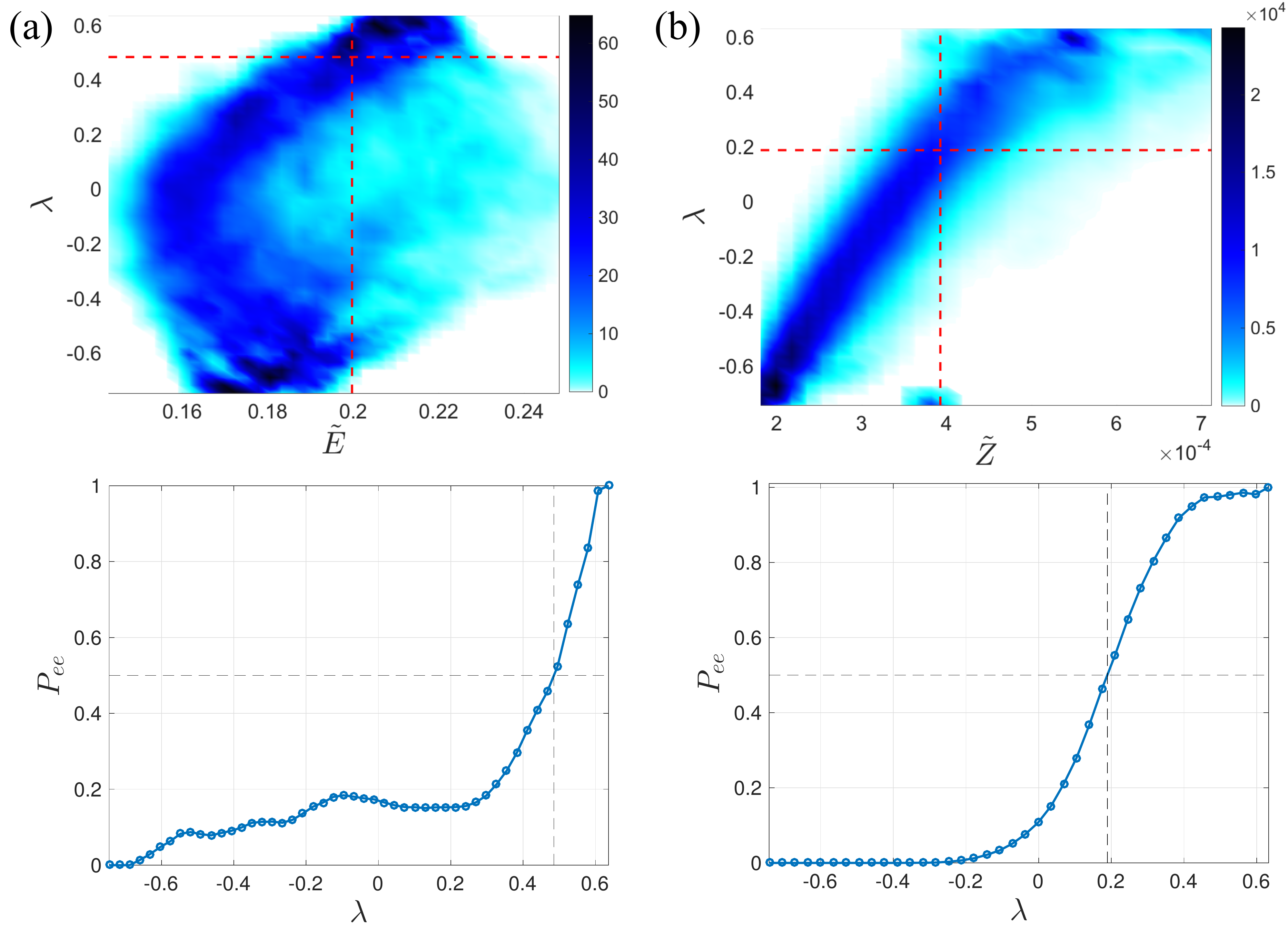}
\caption{(a) Top row: Conditional PDF $p_{\tilde E|\lambda}$ computed with 
$t_0=t_e$ and $\Delta t = 10t_e$ where $t_e$ denotes the eddy turnover time. 
The vertical dashed line marks the threshold $E_e$ of extreme events
that is prescribed as the mean of $E$ plus one standard deviation. 
The horizontal dashed line marks the extreme event threshold $\lambda_e$ according
to the indicator $\lambda$.
Bottom row: The probability of upcoming extreme events $P_{ee}$. The horizontal
dashed line marks $P_{ee}=0.5$ and the vertical dashed line marks the 
extreme event threshold $\lambda_e$ so that $P_{ee}(\lambda_e)=0.5$.
(b) Same as panel (a) but the figure correspond to the energy dissipation $Z$
versus the indicator $\lambda$.
}
\label{fig:R2200_pred_dt1}
\end{figure}

The next step is to introduce a time lag in order to investigate whether
the extreme indicator values proceed the extreme episodes of energy and dissipation. 
This is clearly the case in figure~\ref{f:ee_ke_diss_lam}, however,
we show that this is true most of the time during the longterm simulations. 
Figure~\ref{fig:R2200_pred_dt1} shows the conditional PDFs $p_{\tilde E|\lambda}$
and $p_{\tilde Z|\lambda}$ where the future maxima $\tilde E$ and $\tilde Z$ 
are computed with $t_0=t_e$ and $\Delta t = 10t_e$ (see equation~\eqref{eq:rv_max}).
The extreme value threshold $E_e$ (respectively, $Z_e$) are 
set as the mean of energy (respectively, dissipation) plus one standard deviation.
Figure~\ref{fig:R2200_pred_dt1} also shows the 
corresponding $P_{ee}$ computed from equation~\eqref{eq:Pee}. 
The extreme event threshold according to the indicator $\lambda$ is 
the point at which $P_{ee}(\lambda_e)=0.5$.

\begin{figure}
\centering
\includegraphics[width=.95\textwidth]{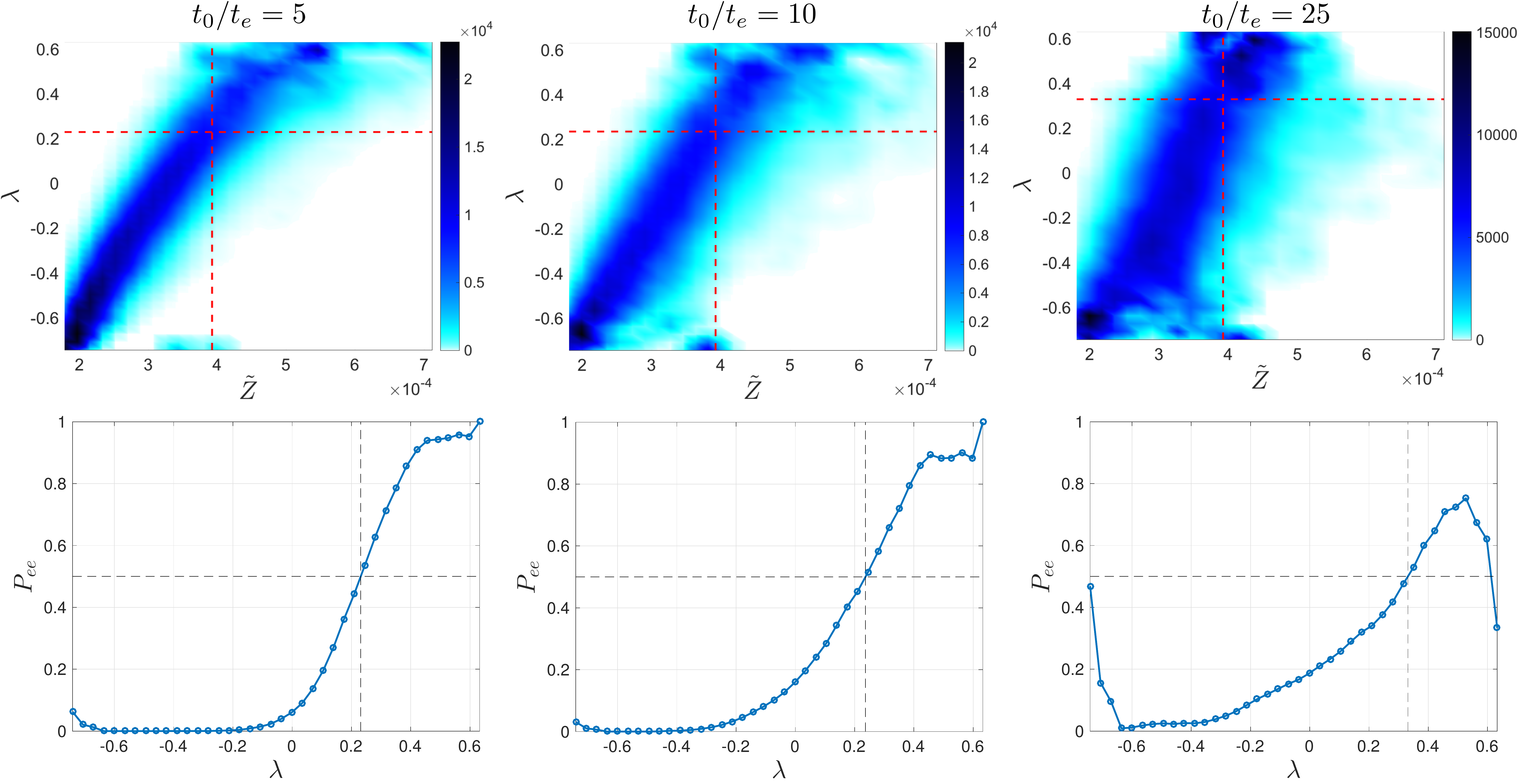}
\caption{Prediction quality for varying prediction times $t_0$. In all plots, the 
length of the future maximum time window in Eq.~\eqref{eq:rv_max} is $\Delta t=10$.
First row shows the conditional PDF $p_{\tilde Z|\lambda}$
and the second row shows the corresponding probability of future extreme events defined in Eq.~\eqref{eq:Pee}.}
\label{fig:R2200_Zlam_dts}
\end{figure}

As the prediction time $t_0$ increases, we expect the prediction skill of the indicator 
to deteriorate. This is shown in figure~\ref{fig:R2200_Zlam_dts} where the 
conditional PDF $p_{\tilde Z|\lambda}$ are shown for increasing prediction times $t_0$. 
For prediction times as large as $t_0=10t_e$ the prediction skill of the indicator 
is still reasonably satisfactory. However, as the prediction time increases to 
close $t_0=25t_e$ the predictor returns significant amount of false positives and 
false negatives, thus losing its predictive value.
The rates of successful predictions and successful rejections, defined in equation~\eqref{eq:rs},
are reported in Table~\ref{tab:rs} for a few prediction time horizons. 

\begin{table}
\centering
\caption{The prediction skill of the indicator $\lambda$ at two Reynolds numbers.
The rate of successful rejections (RSR) and the rate of successful predictions (RSP) are reported for the energy dissipation rate $Z$.
The parameters $t_0$ and $\Delta t$ denote the prediction time parameters
defined in~\eqref{eq:rv_max}.}
\begin{tabular}{|c|c|c|c|c|}
\cline{1-5}
$\vc{Re}$ & $\vc{t_0/t_e}$ & $\vc{\Delta t/t_e}$ & \textbf{RSP} & \textbf{RSR} \\ \hline
\multirow{3}*{$\vc{2200}$} & 1 & 10 & $86.9\%$ & $94.7\%$ \\ \cline{2-5}
                                           & 2 & 10 & $86.7\%$ & $94.3\%$ \\ \cline{2-5}
                                           & 3 & 10 & $86.4\%$ & $93.8\%$ \\ \hline
\multirow{3}*{$\vc{3000}$} & 1 &  5 & $79.4\%$ & $91.7\%$ \\ \cline{2-5}
                                           & 2 &  5 & $75.4\%$ & $92.6\%$ \\ \cline{2-5}
                                           & 3 &  5 & $74.9\%$ & $92.2\%$ \\ \hline
\end{tabular}
\label{tab:rs}
\end{table}

\begin{figure}
\centering
\subfigure[]{\includegraphics[width=.3\textwidth]{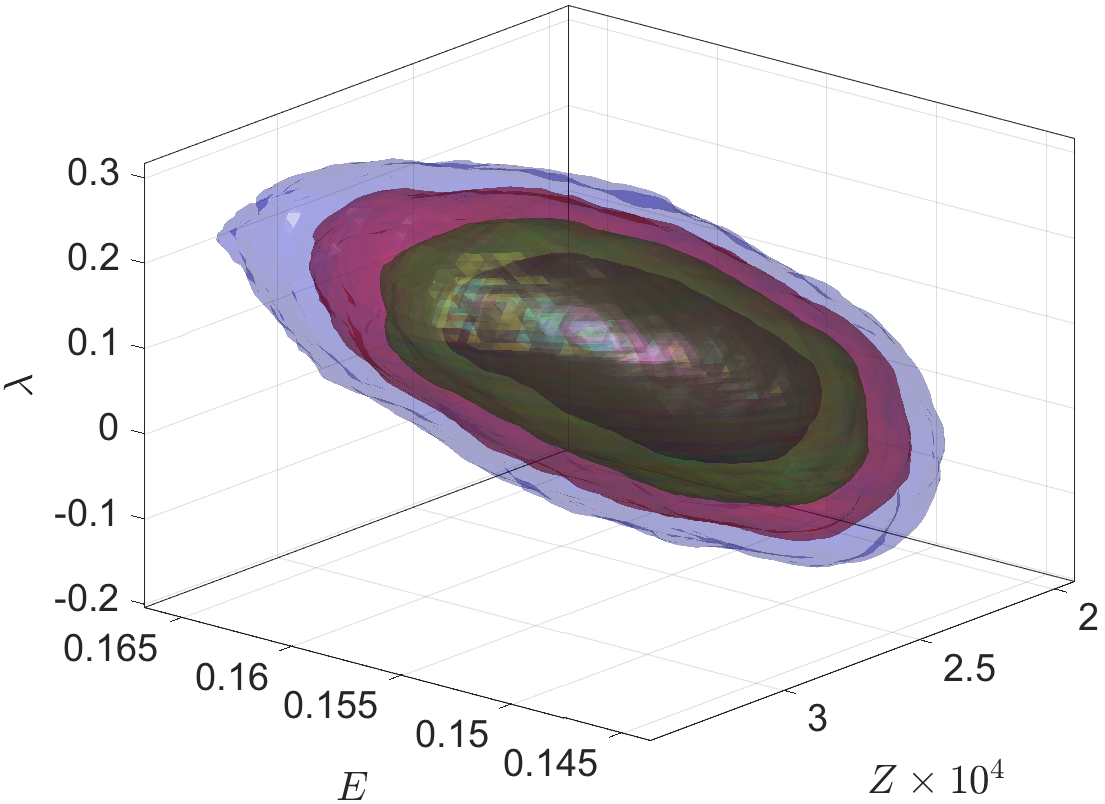}}
\subfigure[]{\includegraphics[width=.3\textwidth]{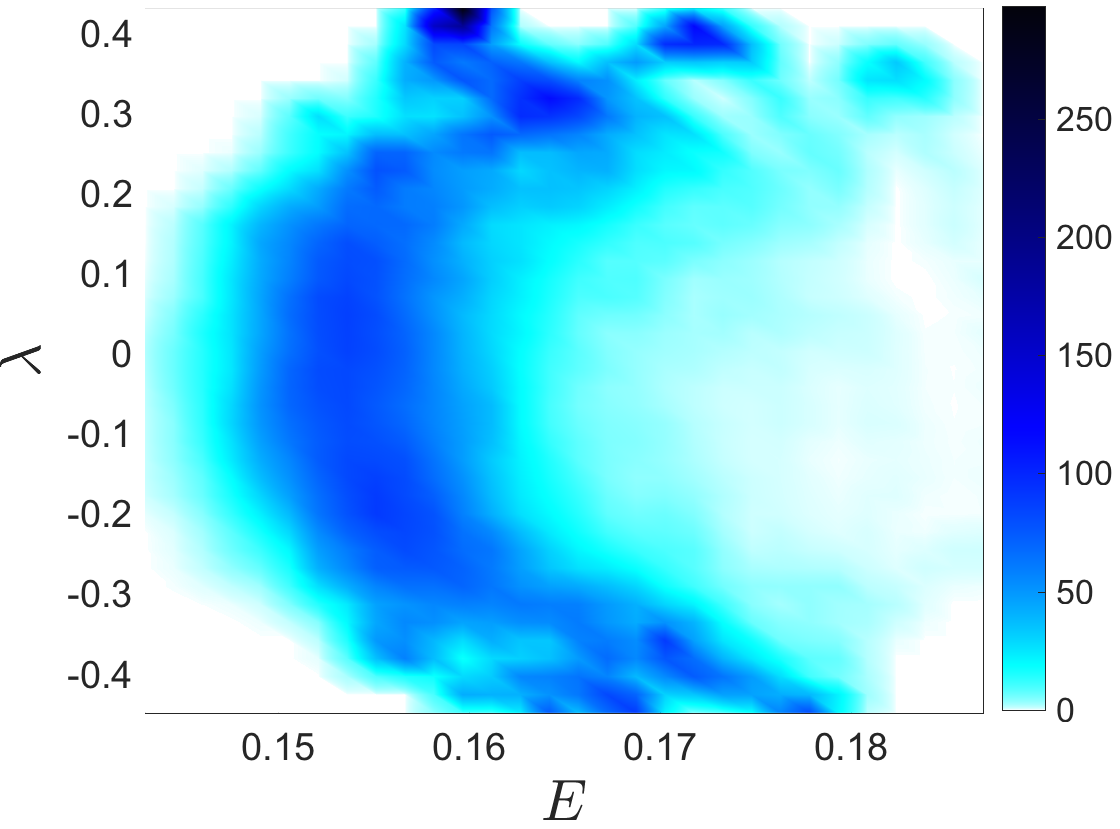}}
\subfigure[]{\includegraphics[width=.3\textwidth]{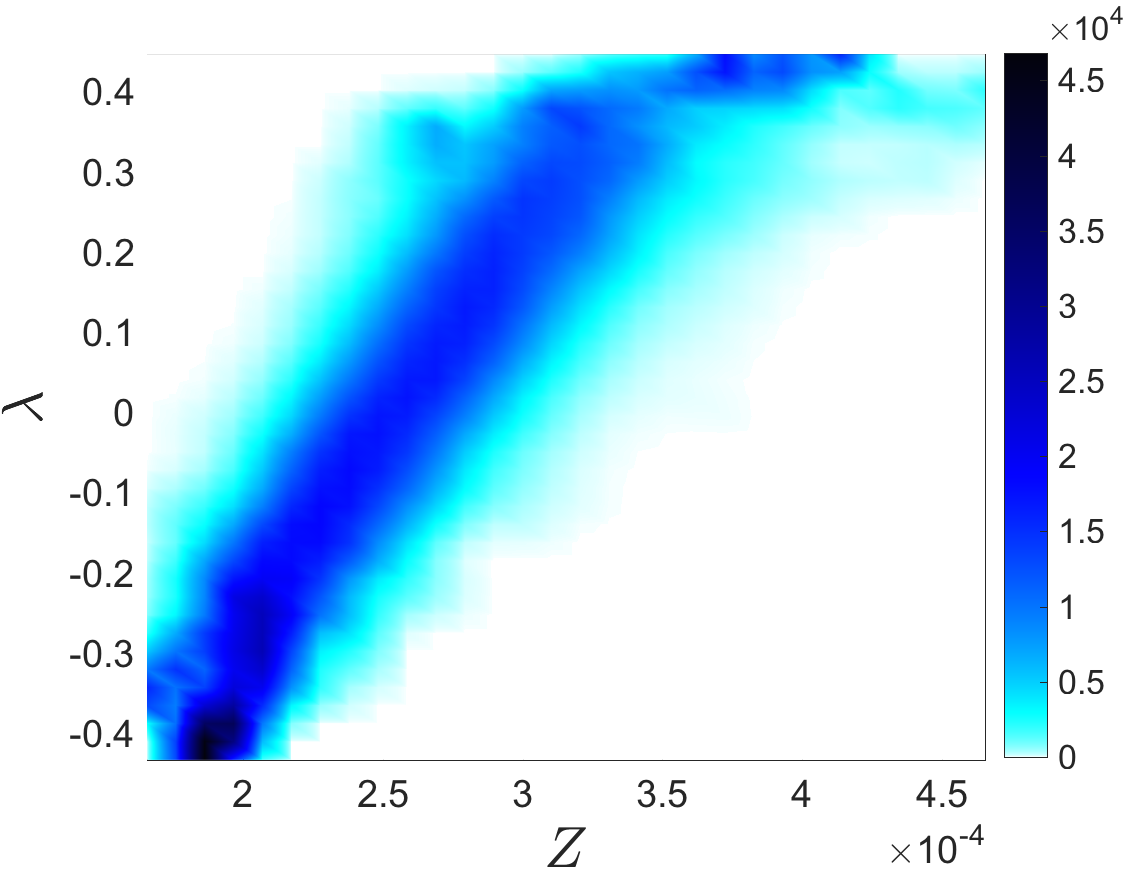}}
\caption{$Re=3000$. 
(a) Joint PDF of the kinetic energy $E$, energy dissipation $Z$ and the indicator $\lambda$.
(b) Conditional PDF of $Y_t=\mbox{Indicator}$ and $X_t=\mbox{Energy}$. 
(c) Conditional PDF of $Y_t=\mbox{Indicator}$ and $X_t=\mbox{Energy Dissipation}$.
}
\label{fig:R3000_dt0}
\end{figure}

Similar results are observed at higher Reynolds numbers. Figure~\ref{fig:R3000_dt0}
 shows the joint and conditional PDFs of the indicator, energy and dissipation at $Re=3000$.
These PDFs resemble those of figure~\ref{fig:R2200_dt0} for the lower Reynold number $Re=2200$. 
This demonstrates clearly the robustness of the derived indicator. 
Table~\ref{tab:rs} also contains the rates of success in predicting extreme (and non-extreme) events
at $Re=3000$.
We point out that the channel flow at some intermediate Reynolds numbers 
between $Re=2200$ and $Re=3000$
did not exhibit extreme events in the time horizons we simulated. 

\section{Conclusions}\label{sec:concl}

We have demonstrated an original approach for the derivation of precursors of extreme events in a challenging problem involving a turbulent channel flow. The extreme events in this case have the form of random near-laminarization episodes that lead to bursts of the kinetic energy and the energy dissipation rate. We formulate a constrained optimization problem that searches for initial states with the most intense growth of kinetic energy, within a constrained set in the core of the underlying turbulent attractor. By searching over a high probability set, we achieve a numerically tractable optimization problem, while at the same time we exclude exotic states that may correspond to intense growth of energy but have very low probability to occur if we are close to the attractor. 

The derived precursor is demonstrated to successfully capture extreme dissipation episodes several eddy turnover times before the event. We have discussed its physical relevance and have demonstrated its robustness as the Reynolds number of the flow changes. Because the developed scheme utilizes full simulations and not linearized approximations, it has the potential to be extended to larger, more complicated flows although divergence of the adjoint might be a limiting factor for obtaining the solution of the associated optimization problem. The success of the presented approach to an intermittently turbulent
channel flow implies the potential of the method for studying transitional flows, such as bypass transition of boundary layers. Our future endeavors include the utilization of these precursors for the control and suppression of extreme events in these systems.  

\subsubsection*{Acknowledgments}
PJB was supported by an appointment to the NASA Postdoctoral Program at the NASA Ames Research Center, administered by Universities Space Research Association under contract with NASA. This research was sponsored by NASA's Transformational Tools and Technologies (TTT) Project of the Transformative Aeronautics Concepts Program under the Aeronautics Research Mission Directorate. TPS and MF have been supported through the ARO MURI W911NF-17-1-0306 and the ONR grant N00014-15-1-2381.
Resources supporting this work were provided by the NASA High-End Computing (HEC) Program through the NASA Advanced Supercomputing (NAS) Division at Ames Research Center.


\end{document}